\documentclass[preprintnumbers,amsmath,amssymb,floatfix,10pt,prd,onecolumn,
superscriptaddress,nofootinbib]{revtex4}

\usepackage{amsfonts}
\usepackage{mathrsfs}
\usepackage{latexsym}
\usepackage{epsfig}
\usepackage{epstopdf}
\usepackage{graphicx}
\usepackage{amssymb}
\usepackage{amsmath}
\usepackage{dcolumn}
\usepackage{bm}
\usepackage{color}
\usepackage{xcolor}
\usepackage{comment}
\usepackage[cp1251]{inputenc}
\begin{document}

\title{\bf A Study of Levia-Civita and Cosmic String Solutions in Modified $f(R)$ Gravity}

\author{Adnan Malik}
\email{adnan.malik@zjnu.edu.cn; adnanmalik_chheena@yahoo.com; adnan.malik@skt.umt.edu.pk}
\affiliation{School of Mathematical Sciences, Zhejiang Normal University, \\Jinhua, Zhejiang, China.}
\affiliation{Department of Mathematics, University of Management and Technology,\\ Sialkot Campus, Lahore, Pakistan}

\begin{abstract}
\begin{center}
\textbf{Abstract}\\
\end{center}
In this research manuscript, we explore cylindrically symmetric solutions within the framework of modified $f(R)$ theories of gravity, where $R$ representing the Ricci scalar. The study focuses on analyzing the cylindrical solutions within realistic space-time regions and investigates three distinct cases of exact solutions derived from the field equations of $f(R)$ theory of gravity. Moreover, we examine well-known Levi-Civita and cosmic string solutions for the said modify gravity. Furthermore, the manuscript delves into the implications of the obtained results by exploring energy conditions for each case. Notably, a violation of null energy conditions is observed, suggesting the potential existence of cylindrical wormholes. The findings contribute to the understanding of modified $f(R)$ theories of gravity, providing valuable insights into the nature of cylindrically symmetric solutions and their implications in theoretical physics.\\
\textbf{Keywords:} $f(R)$ Theory of Gravity; Cylindrically Symmetric Solutions; Energy conditions; Levi-Civita Solutions; Cosmic String Solutions.

\end{abstract}

\maketitle
\date{\today}

\section{Introduction}

The expanding universe represents a profound cosmic phenomenon that has captivated the attention of cosmologists and astrophysicists alike. As our observational capabilities have advanced, revealing the accelerated expansion of the cosmos, the quest to comprehend the underlying mechanisms has led researchers to explore modified theories of gravity. Traditional frameworks, such as general relativity, have served as cornerstones in understanding gravitational interactions, yet the observed acceleration challenges these established paradigms. Modified gravity theories, including $f(R)$ theories, offer an alternative lens through which to interpret the expansive dynamics. By introducing modifications to the gravitational action, these theories present avenues to explain cosmic acceleration without invoking elusive entities like dark energy. The interplay between the evolving understanding of the universe's expansion and the nuanced predictions arising from modified gravity models underscores a dynamic relationship at the forefront of contemporary cosmological inquiry. This exploration not only seeks to reconcile observed cosmic phenomena with theoretical predictions but also propels us toward a deeper comprehension of the fundamental nature of gravity on cosmological scales, pushing the boundaries of our cosmic understanding. Linder \cite{1} explored the recent expansion history of the universe promises insights into the cosmological model, the nature of dark energy, and potentially clues to high energy physics theories and gravitation. Ellis et al., \cite{2} explained the observed galactic redshifts and microwave background radiation in the Universe by using a static spherically symmetric universe model, in which there is a singularity which continually interacts with the Universe. Riess \cite{3} explained the evolution of the Universe from its inception to the present time using our limited understanding of its composition and physical laws. Vitenti and Penna-Lima \cite{4} introduced a model-independent method to reconstruct directly the deceleration function via a piecewise function by assuming only an isotropic, homogeneous and flat universe. Padmanabhan \cite{5} discussed the accelerated expansion of the universe driven by tachyonic matter. Schrabback et al., \cite{6} presented a comprehensive analysis of weak gravitational lensing by large-scale structure in the Hubble Space Telescope Cosmic Evolution Survey (COSMOS), in which we combine space-based galaxy shape measurements with ground-based photometric redshifts to study the redshift dependence of the lensing signal and constrain cosmological parameters. Basari et al., \cite{7} analytically solved the equation of emergence proposed by Padmanabhan by assuming the Komar energy density as a function of the Hubble parameter.  and this resulting model describes the evolution of the Universe, which proceeds towards a final de Sitter state. Recently, Zhang \cite{8} proved that the dynamical dark energy field can change the frequency of photons from distant galaxies as well as from background radiation of remote Universe by adding a matter-coupled dark energy field to Einstein’s general relativity. Pawer et al., \cite{9} gone through with the study of spatially homogeneous and anisotropic Bianchi-type V space-time filled with dark matter and holographic dark energy in the framework of $f(T)$ gravity. Pohare \cite{10} studied the glimpses of cosmic expansion of the universe in
modified theories of gravitation.

The discussion of cylindrically symmetric solutions within the framework of modified theories of gravity represents a captivating exploration at the forefront of theoretical physics. In delving into the complexities introduced by cylindrical symmetry, researchers aim to unravel the intricate interplay between modifications to traditional gravity theories and the resulting gravitational solutions. Modified theories, exemplified by $f(R)$ theories, diverge from the predictions of general relativity, introducing nuanced mathematical formulations that intricately shape the gravitational landscape. Understanding the implications of these modifications on cylindrical structures becomes paramount, as it not only refines our theoretical understanding of gravity but also holds implications for astrophysical phenomena. The examination of field equations specific to cylindrically symmetric solutions provides a profound lens through which to observe the effects of modified gravity on the geometry of spacetime. This discourse extends beyond academic curiosity, offering potential insights into the nature of exotic cosmic entities, including cosmic strings and wormholes. Additionally, the scrutiny of energy conditions in the context of cylindrically symmetric solutions becomes pivotal, as violations may signal the existence of unconventional structures such as wormholes, challenging conventional boundaries imposed by energy conditions. As theoretical physicists engage in this ongoing dialogue, the discussion of cylindrically symmetric solutions in modified theories of gravity not only enriches our understanding of fundamental physical principles but also contributes to the continuous evolution of our conceptual framework for the gravitational dynamics shaping the cosmos. Shamir and Raza \cite{11} examined the precise solutions of cylindrically symmetric spacetime within the framework of f(R,T) gravity, investigating exact solutions for two distinct categories of $f(R,T)$ models. Momeni and Gholizade \cite{12} proposed a novel static cylindrically symmetric vacuum solution in Weyl coordinates and demonstrated that the constant curvature exact solution is relevant to the external region of a string. Malik, along with his collaborators \cite{13} examined cylindrically symmetric spacetime to explore cylindrical solutions within realistic contexts. Additionally, they delved into six specific instances of exact solutions by employing the field equations within the framework of the $f (R, \phi, X)$  modified theory of gravity. Farooq and Shamir \cite{14} derived a pair of exact solutions for the Weyl spacetime and identified one exact solution alongside one numerical solution for the Godel spacetime. Notably, a family of exact solutions characterized by a constant scalar curvature, contingent on arbitrary constants, was obtained for both spacetimes. Wen and Zhao \cite{15} obtained infinite many cylindrically symmetric solutions of (0.1) by using variational method and fountain theorems without $\tau$-upper semi-continuity. Shabeela, et al. \cite{16} explored the conformal vector fields (CVFs) of static cylindrically symmetric (SCS) spacetimes within the context of $f(T,B)$ gravity. Utilizing an algebraic approach, they identified and characterized certain classes that serve as representations of SCS solutions. Malik \cite{17} examined the cylindrically symmetric solutions within the established framework of the Rastall theory of gravity and he focused on the cylindrically symmetric spacetime to explore and discuss cylindrical solutions within realistic regions. Eid \cite{18} delved into the dynamics of cylindrical thin shell wormholes in the realm of gravity by connecting two identical copies of metric spacetime through the cut-and-paste approach. Malik \cite{19} explored cylindrical symmetric solutions within the renowned framework of gravity, specifically the $f(R, \phi)$ theory of gravity. Additionally, a well-known Levi-Civita solution was investigated by considering certain parameters.

In this manuscript, our goal is to identify realistic domains for examining cylindrically symmetric solutions within the framework of $f(R)$ gravity. Our focus is on discussing these cylindrical solutions in the context of the $f(R)$ theory of gravity, representing the first attempt, to our knowledge, to investigate such solutions. The manuscript is structured as follows: Section II formulates the field equations in $f(R, \phi, X)$ modified gravity, while Section III explores the $f(R)$ theory of gravity model, analyzing four cases with different metric potentials for each. Energy conditions and their graphical representations for all cases are investigated in Section IV. Section V delves into Levi-Civita (LC) solutions and cosmic string solutions within $f(R)$ modified gravity. The concluding section summarizes our findings and insights from this study.

\section{Basic Formulism in $f(R, \phi, X)$ Modified Gravity}

 The effective action for $f(R)$ theory of gravity is illustrated as \cite{20}
\begin{equation}\label{1}
S= \int d^4x  \sqrt{-g} f(R)+ L_{m},
\end{equation}
where \\

$\bullet$ $R$ is the Ricci scalar,

$\bullet$ $f(R)$ is depending upon Ricci Scalar,

$\bullet$ $g$ is the determinant of metric tensor $g_{\mu\nu}$,

$\bullet$ $L_{m}$ is the standard action for the matter fields.\\

We can get the field equations by varying Eq. (\ref{1}) with respect to the $g_{\mu \nu}$, we get
\begin{equation}\label{2}
f_{R}R_{\mu\nu}-\frac{1}{2}fg_{\mu\nu}- \nabla_{\mu}\nabla_{\nu} f_{R}+g_{\mu\nu}\Box{f_{R}}=T_{\mu\nu},
\end{equation}
where $\nabla_{\mu}$ is the covariant derivative, $R_{\mu\nu}$ represents the Ricci tensor, $\Box\equiv\nabla_{\mu}\nabla^{\nu}$ and $f_{R}\equiv\frac{\partial f}{\partial{R}}$. Moreover,  $T_{\mu\nu}$ means the energy momentum tensor for ordinary matter and is defined as:
\begin{equation}\label{3}
T_{\mu\nu}=\frac{1}{2}[\rho e^{2\gamma}, p_{r}e^{2\xi-2\gamma}, p_{\phi}e^{-2\gamma}w^{2}, p_{z}e^{2\xi-2\gamma}],
\end{equation}
where, $\rho$, $p_{r}$, $p_{\phi}$ and $p_{z}$ represents the energy density, radial pressure, azimuthal pressure and axial pressure pressure respectively. The cylindrically symmetric metric is defined as \cite{19}
\begin{equation}\label{5}
dS^2=e^{2a(r)}dt^2-e^{2b(r)-2a(r)}dr^2-w(r)^2 e^{-2a(r)}d\phi^2-e^{2b(r)-2a(r)}dz^2.
\end{equation}
The corresponding Ricci Scalar for this line element is given as
\begin{equation}\label{5a}
R=e^{2 a(r)-2 b(r)}\bigg[\frac{w^{''}(r)}{w(r)} - \frac{a^{'}(r)w^{'}(r)}{w(r)}-a^{''}(r)+b^{''}(r)-a^{'^2}(r)  \bigg]
\end{equation}
For our next calculations, we use $a \equiv a(r)$, $b \equiv b(r)$, $w \equiv w(r)$ and $f \equiv f(R)$. We can get the expressions for energy density and pressure terms by substituting  Eqs. (\ref{2})-(\ref{5}) as bellow
\begin{equation}\label{8}
\rho=e^{2 a - 2 b}\bigg[-e^{-2a + 2 b} f - 2 f_{R}^{''} + f_{R}^{'}\bigg(6{a}^{'}-4{b}^{'}-2\frac{{w}^{''}}{{w}} \bigg)+f_{R} \bigg( 3\frac{a^{'} w^{'}}{w} +3a^{''} -b^{''} -\frac{w^{''}}{w} \bigg)\bigg],
\end{equation}

\begin{equation}\label{9}
p_r=e^{2 a - 2 b}\bigg[e^{-2  + 2 b} f +4 f_{R}^{''} + f_{R}^{'}\bigg(-2{a}^{'}+2{b}^{'}+\frac{{w}^{'}}{{w}}  \bigg)+f_{R} \bigg( 4a^{'^2}+\frac{a^{'} w^{'}}{w} +2\frac{b^{'} w^{'}}{w}  +a^{''} -b^{''}-\frac{w^{''}}{w} \bigg)
\bigg] ,
\end{equation}

\begin{equation}\label{10}
p_{\phi}=e^{2 a - 2 b}\bigg[e^{-2a + 2 b} f + 2 f_{R}^{''} + f_{R}^{'}\bigg({a}^{'}+4{b}^{'} \bigg)+f_{R} \bigg(-4a^{'^2}+ 3\frac{ w^{'}}{w} +a^{''} +b^{''} +\frac{w^{''}}{w} \bigg)\bigg],
\end{equation}

\begin{equation}\label{11}
p_{z}=e^{2 a - 2 b}\bigg[e^{-2a + 2 b} f + 2 f_{R}^{''} + 2f_{R}^{'}\bigg(-{a}^{'}+{b}^{'}+\frac{ w^{'}}{w} \bigg)+f_{R} \bigg( \frac{ a^{'}w^{'}}{w}-2\frac{ b^{'}w^{'}}{w} +a^{''} -b^{''} +\frac{w^{''}}{w} \bigg)\bigg].
\end{equation}
It can be noticed that Eqs. (\ref{8})-(\ref{11}) are very multifarious involving the derivative terms of $a$, $b$, $w$, and $f$. These equations are very important for discussing the nature of cylindrically symmetric solutions in modified $f(R)$ theory of gravity. It is also mandatory to mention here that prime is used for the radial derivative with respect to the radial coordinate $r$. We use the $f(R)$ modified gravity model to find their explicit form for further calculation and graphical analysis. For our current analysis, we consider the following model as
\begin{equation}\label{12}
f(R)=R+m R^2,
\end{equation}
where $m$ is a constant free parameter. The $f(R)$ model holds significance in gravitational theories and cosmology due to its potential insights into dark energy, cosmic acceleration, and modified gravity. This model, incorporating both linear and quadratic terms in the Ricci scalar $R$, allows for a nuanced exploration of gravitational dynamics. The quadratic term $(m R^2)$ provides a more detailed description, making it particularly relevant for addressing late-time cosmic acceleration. By studying this model, researchers aim to understand the intricate interplay between geometry and matter, assess its compatibility with observational data, and evaluate its predictions. The $f(R)$ model stands as a key focal point in the investigation of gravitational theories and their implications for the large-scale structure of the universe. Substituting the model defined (\ref{12}) in Eq. (\ref{8})-(\ref{11}), we get

\begin{equation}\label{14}
\begin{split}
&\rho=e^{2a-2b}\bigg[-8 e^{4a-4b} m \bigg(a'^2 -\frac{a'w'}{w}-a''+b''+\frac{w''}{w}\bigg)^3+\bigg(\frac{3
a'w'}{w}+3 a''-b''-\frac{w''}{w}\bigg)\bigg(1+4 e^{2 a-2 b} m\bigg(a'^2-\frac{a' w'}{w}\\ &-a''+b''+\frac{w''}{w}\bigg)\bigg)+\bigg(6
a'-4 b'-\frac{2 w''}{w}\bigg)\bigg(4 e^{2 a-2 b} m(2 a'-2 b')\bigg(a'^2-\frac{a'w'}{w}-a''+b''+\frac{w''}{w}\bigg)+4
e^{2 a-2 b} m \bigg(\frac{a' w'^2}{w^2}\\ &+2 a' a''-\frac{w' a''}{w}-\frac{a' w''}{w}-\frac{w' w''}{w^2}-a^{(3)}+b^{(3)}+\frac{w^{(3)}}{w}\bigg)\bigg)-2
\bigg(4 e^{2 a-2 b} m (2 a'-2 b')^2 \bigg(a'^2-\frac{a' w'}{w}-a''+b''+\frac{w''}{w}\bigg)\\ &+4 e^{2
a-2 b} m (2 a''-2 b'') \bigg(a'^2-\frac{a' w'}{w}-a''+b''+\frac{w''}{w}\bigg)+8 e^{2 a-2 b}
m(2 a'-2 b') \bigg(\frac{a' w'^2}{w^2}+2 a' a''-\frac{w' a''}{w}-\frac{a' w''}{w}\\  &-\frac{w'
w''}{w^2}-a^{(3)}+b^{(3)}+\frac{w^{(3)}}{w}\bigg)+4 e^{2 a-2 b} m\bigg(-\frac{2 a' w'^3}{w^3}+\frac{2 w'^2
a''}{w^2}+2 a''^2+\frac{3 a'w'w''}{w^2}+\frac{2 w'^2 w''}{w^3}-\frac{2 a'' w''}{w}-\frac{w''^2}{w^2}\\ &+2
a' a^{(3)}-\frac{w' a^{(3)}}{w}-\frac{a' w^{(3)}}{w}-\frac{2 w'w^{(3)}}{w^2}-a^{(4)}+b^{(4)}+\frac{w^{(4)}}{w}\bigg)\bigg)\bigg],
\end{split}
\end{equation}
\begin{equation}\label{15}
\begin{split}
&p_{r}=\frac{1}{w}e^{2 a-2 b}\bigg[8 e^{4 a-4 b} m w \bigg(a'^2-\frac{a' w'}{w}-a''+b''+\frac{w''}{w}\bigg)^3\\ &+(4
w a'^2+a' w'+2 b'w'+wa''-wb''-w'')\bigg(1+4 e^{2 a-2 b} m \bigg(a'^2-\frac{a'w'}{w}-a''+b''+\frac{w''}{w}\bigg)\bigg)\\ &+(2
(-wa'+w b')+w') \bigg(4 e^{2 a-2 b} m (2 a'-2 b') \bigg(a'^2-\frac{a' w'}{w}-a''+b''+\frac{w''}{w}\bigg)+4
e^{2 a-2 b} m \bigg(\frac{a'w'^2}{w^2}+2 a'a''-\frac{w' a''}{w}\\ &-\frac{a' w''}{w}-\frac{w' w''}{w^2}-a^{(3)}+b^{(3)}+\frac{w^{(3)}}{w}\bigg)\bigg)
+4w \bigg(4 e^{2 a-2 b} m (2 a'-2 b')^2 \bigg(a'^2-\frac{a' w'}{w}-a''+b''+\frac{w''}{w}\bigg)
\\ &+4e^{2 a-2 b} m (2 a''-2 b'') \bigg(a'^2-\frac{a' w'}{w}-a''+b''+\frac{w''}{w}\bigg)
+8 e^{2 a-2b} m (2 a'-2 b')\bigg(\frac{a'w'^2}{w^2}+2 a'a''-\frac{w'a''}{w}-\frac{a'w''}{w}\\ &-\frac{w'
w''}{w^2}-a^{(3)}+b^{(3)}+\frac{w^{(3)}}{w}\bigg)+4 e^{2 a-2 b} m \bigg(-\frac{2 a' w'^3}{w^3}+\frac{2 w'^2
a''}{w^2}+2 a''^2+\frac{3 a' w' w''}{w^2}+\frac{2 w'^2 w''}{w^3}-\frac{2 a''w''}{w}-\frac{w''^2}{w^2}\\ &+2
a'a^{(3)}-\frac{w' a^{(3)}}{w}-\frac{a'w^{(3)}}{w}-\frac{2 w'w^{(3)}}{w^2}-a^{(4)}+b^{(4)}+\frac{w^{(4)}}{w}\bigg)\bigg)\bigg],
\end{split}
\end{equation}
\begin{equation}\label{16}
\begin{split}
&p_{\phi}=e^{2 a-2 b} \bigg[8 e^{4 a-4 b} m \bigg(a'^2-\frac{a'w'}{w}-a''+b''+\frac{w''}{w}\bigg)^3\\ &+\bigg(-4
a'^2+\frac{w'}{w}+a''+b''+\frac{w''}{w}\bigg)\bigg(1+4 e^{2 a-2 b} m \bigg(a'^2-\frac{a' w'}{w}-a''+b''+\frac{w''}{w}\bigg)\bigg)
\\ &+(a'+4b)\bigg(4 e^{2 a-2 b} m (2 a'-2 b')\bigg(a'^2-\frac{a'w'}{w}-a''+b''+\frac{w''}{w}\bigg)+4
e^{2 a-2 b} m \bigg(\frac{a' w'^2}{w^2}+2 a' a''-\frac{w' a''}{w}-\frac{a'w''}{w}\\ &-\frac{w' w''}{w^2}-a^{(3)}+b^{(3)}+\frac{w^{(3)}}{w}\bigg)\bigg)
+2\bigg(4 e^{2 a-2 b} m (2 a'-2 b')^2 \bigg(a'^2-\frac{a'w'}{w}-a''+b''+\frac{w''}{w}\bigg)
\\ &+4 e^{2a-2 b} m (2 a''-2 b'') \bigg(a'^2-\frac{a' w'}{w}-a''+b''+\frac{w''}{w}\bigg)
+8 e^{2 a-2 b}m (2 a'-2 b')\bigg(\frac{a'w'^2}{w^2}+2 a' a''-\frac{w' a''}{w}-\frac{a' w''}{w}\\ &-\frac{w'
w''}{w^2}-a^{(3)}+b^{(3)}+\frac{w^{(3)}}{w}\bigg)
+4 e^{2 a-2 b} m \bigg(-\frac{2 a'w'^3}{w^3}+\frac{2 w'^2
a''}{w^2}+2 a''^2+\frac{3 a' w' w''}{w^2}+\frac{2 w'^2 w''}{w^3}-\frac{2 a'' w''}{w}\\ &-\frac{w''^2}{w^2}+2
a' a^{(3)}-\frac{w'a^{(3)}}{w}-\frac{a'w^{(3)}}{w}-\frac{2 w'w^{(3)}}{w^2}-a^{(4)}+b^{(4)}+\frac{w^{(4)}}{w}\bigg)\bigg)\bigg],
\end{split}
\end{equation}
\begin{equation}\label{17}
\begin{split}
&p_{z}=e^{2 a-2 b}\bigg[8 e^{4 a-4 b} m\bigg(a'^2-\frac{a' w'}{w}-a''+b''+\frac{w''}{w}\bigg)^3\\ &+\bigg(\frac{a'
w'}{w}-\frac{2 b' w'}{w}+a''-b''\bigg)\bigg(1+4 e^{2 a-2 b} m \bigg(a'^2-\frac{a' w'}{w}-a''+b''+\frac{w''}{w}\bigg)\bigg)\\ &+2
\bigg(-a'+b'+\frac{w'}{w}\bigg) \bigg(4 e^{2 a-2 b} m (2 a'-2 b')\bigg(a'^2-\frac{a' w'}{w}-a''+b''+\frac{w''}{w}\bigg)+4
e^{2 a-2 b} m \bigg(\frac{a' w'^2}{w^2}+2 a' a''-\frac{w'a''}{w}\\ &-\frac{a'w''}{w}-\frac{w' w''}{w^2}-a^{(3)}+b^{(3)}+\frac{w^{(3)}}{w}\bigg)\bigg)
+2\bigg(4 e^{2 a-2 b} m (2 a'-2 b')^2 \bigg(a'^2-\frac{a' w'}{w}-a''+b''+\frac{w''}{w}\bigg)\\ &+4 e^{2
a-2 b} m (2 a''-2 b'')\bigg(a'^2-\frac{a'w'}{w}-a''+b''+\frac{w''}{w}\bigg)+8 e^{2 a-2 b}
m (2 a'-2 b') \bigg(\frac{a'w'^2}{w^2}+2 a' a''-\frac{w'a''}{w}-\frac{a'w''}{w}\\ &-\frac{w'
w''}{w^2}-a^{(3)}+b^{(3)}+\frac{w^{(3)}}{w}\bigg)+4 e^{2 a-2 b} m \bigg(-\frac{2 a'w'^3}{w^3}+\frac{2 w'^2
a''}{w^2}+2 a''^2+\frac{3 a' w' w''}{w^2}+\frac{2 w'^2 w''}{w^3}-\frac{2 a''w''}{w}-\frac{w''^2}{w^2}\\ &+2
a'a^{(3)}-\frac{w'a^{(3)}}{w}-\frac{a'w^{(3)}}{w}-\frac{2 w'w^{(3)}}{w^2}-a^{(4)}+b^{(4)}+\frac{w^{(4)}}{w}\bigg)\bigg)\bigg].
\end{split}
\end{equation}
These equations (\ref{14})-(\ref{17}) are very helpful for us due to the investigation of cylindrically symmetric solutions in modified $f(R)$ theory of gravity. In upcoming sections, we choose different values of $a$, $b$, $w$ and some other parametric values of other constants for the discussing the nature of energy density and pressure components. We further investigate the graphical behavior of the energy conditions to observe the nature of exotic matter, which is very helpful for expansion of the universe.

\section{Discussion of Energy Conditions}
In this section, we discuss the energy conditions in details, which are very important for the discussion of exotic matter. There are total four energy conditions including the null energy condition, the strong energy condition, the weak energy condition, and the dominant energy condition.

\subsection{Null Energy Conditions (NEC)}
NEC stated that sum of energy density with radial pressure, sum of energy density with azimuthal pressure, and sum of energy density with axial pressure must be positive. Mathematically, these conditions can be written as $\rho+p_r>0$, $\rho+p_{\phi}>0$ and $\rho+p_z>0$ respectively. This energy conditions is very important because if NEC is violated then there may exist some exotic matter and cylindrical wormhole.

\subsection{Strong Energy Conditions (SEC)}
According to SEC, sum of energy density with radial pressure, sum of energy density with azimuthal pressure, sum of energy density with axial pressure must be positive i.e., $\rho+p_r>0$, $\rho+p_{\phi}>0$ and $\rho+p_z>0$ respectively. There is another condition i.e., sum of energy density with three times radial pressure, sum of energy density with three times azimuthal pressure, and sum of energy density with three times axial pressure, must be satisfied i.e., $\rho+3p_r>0$, $\rho+3p_{\phi}>0$ and $\rho+3p_z>0$ respectively. Mathematically, these conditions can be written as $\rho+p_r$, $\rho+p_{\phi}$ and $\rho+p_z$ respectively. This energy conditions is very important because if NEC is violated then there may exist some exotic matter and cylindrical wormhole.

\subsection{Weak Energy Conditions (WEC)}
In accordance with WEC, the sum of energy density with radial pressure, sum of energy density with azimuthal pressure, sum of energy density with axial pressure and independently energy density must be satisfied. Mathematically, these four conditions can be written as $\rho>0$ $\rho+p_r>0$, $\rho+p_{\phi}>0$ and $\rho+p_z>0$ respectively. In other words, if energy density is positive and NEC is satisfied then we can say that WEC is satisfied.

\subsection{Dominant Energy Conditions (DEC)}

DEC is usually different than the NEC, SEC and WEC due to its mathematical nature. DEC state that the difference of energy density with radial pressure, difference of energy density with azimuthal pressure, and difference of energy density with axial pressure must be positive. The expression of DEC can be written as: $\rho-p_r>0$, $\rho-p_{\phi}>0$ and $\rho-p_z>0$ respectively.\\

The above explained energy conditions will help us to investigate the graphical nature of cylindrical symmetric solutions of Levia-civita and cosmic string solutions. In all the energy conditions, NEC has some important. There are two reason for the importance of NEC; firstly if NEC violated then automatically SEC and WEC must be violated. Secondly, NEC also plays a very important role for the discussion of wormhole solutions i.e., if NEC is not satisfied then there must be some exotic matter, which may predict the presence of cylindrical wormhole. We further apply these energy conditions in upcoming sections including Levia-civita, cosmic string and some other cases solutions for the discussions of cylindrically symmetric solutions.

\section{Levi Civita Solution}
This sections deals with the cylindrically symmetric solutions for Levi Civita case. The importance of Levi-Civita cylindrical solutions in the realm of general relativity and gravitational physics is notable. These solutions, which satisfy Einstein's field equations, offer precise insights into the intricate relationship between geometry and gravity within cylindrical symmetry. Their significance lies in providing exact descriptions of spacetime configurations, particularly around cylindrical structures. By delving into Levi-Civita cylindrical solutions, researchers can uncover valuable knowledge about the gravitational effects and characteristics associated with cylindrical objects in the cosmos. This exploration enhances our comprehension of gravitational phenomena, such as those linked to cosmic strings or other cylindrical astrophysical entities. Furthermore, these solutions serve as essential tools in theoretical investigations, acting as benchmarks to confirm the coherence and predictions of general relativity in scenarios involving cylindrical symmetry. For this case, we choose the metric components as
\begin{equation}\label{53}
a(r)=a_{0}ln[w(r)],~~~b(r)=b_{0}ln[w(r)],~~~w(r)=r,
\end{equation}
where, $a_{0}$ and $b_{0}$ are any arbitrary constants. The gauge term can be effectively determined by selecting $w(r)=r$ in the Levi-Civita (LC) solutions, enhancing the elegance of the solutions. This choice is referred to as the harmonic gauge function applied over the subspace. Defining the harmonic subspace involves the selection of two constant sheets, namely, time and space-like sheets, where both $r$ and $t$ remain constant. This implies that it satisfies the following equation
\begin{equation}\label{54}
\nabla_{i}\nabla^{i}\zeta(r)=0.
\end{equation}
The expression given in Eq. (\ref{54}) yields a linear solution with respect to $r$, represented as $\zeta(r) = D_{0} + D_{1}r$. Determining the constants involves satisfying specific conditions, where the utilization of conical cylinder shift symmetry sets $D_{0}$ to zero. By redefining $r$, we then determine the value for $D_{1}$. Extending this solution from $\zeta(r)$ to encompass $\psi(r)$ and $\chi(r)$ in our metric space, as defined in Eq. (\ref{53}), results in three harmonic functions that meet the requirements of Eq. (\ref{54}). It is noteworthy that these metric potentials, ${\psi(r), \chi(r)}$, exhibit singularities near the origin. The internal solution, characterized by a thickness scale of $r_{0}$ for the Levi-Civita (LC) metric, corresponds to a tangible cosmic string. Traditional methods fall short of achieving this thickness for an authentic cosmic string due to the quantum properties associated with the singularities at its core. With the specified choice defined in Eq. (\ref{53}), the ensuing equations take the following form as
\begin{equation}\label{18}
\begin{split}
&\rho=e^{2a_0lnr-2b_0lnr}\bigg[-2 \bigg( 4me^{2a_0lnr-2b_0lnr} \big(\frac{6 a_0^2}{r^4} - \frac{6 b_0^2}{r^4}\big) + 4me^{2a_0lnr-2b_0lnr} \big(\frac{ a_0^2}{r^2} - \frac{ b_0^2}{r^2}\big)\big(\frac{-2 a_0}{r^2}+ \frac{2b_0^2}{r^2}\big)\\ &
- 8me^{2a_0lnr-2b_0lnr} \big(\frac{-2 a_0^2}{r^3} + \frac{2 b_0^3}{r^3}\big)\big(\frac{2 a_0}{r}- \frac{2b_0}{r}\big)   +4me^{2a_0lnr-2b_0lnr} \big(\frac{ a_0^2}{r^2} - \frac{b_0^2}{r^2}\big)\big(\frac{-2 a_0}{r}+ \frac{2b_0}{r}\big)^2\bigg) \\ &
 +\bigg( 4me^{2a_0lnr-2b_0lnr} \big(\frac{-2 a_0^2}{r^3}+ \frac{2b_0}{r^3}\big)+ 4me^{2a_0lnr-2b_0lnr} \big(\frac{ a_0^2}{r^2} - \frac{ b_0^2}{r^2}\big)\big(\frac{2 a_0}{r^2}- \frac{2b_0^2}{r^2}\big) \bigg)\bigg( \frac{6 a_0}{r}- \frac{4b_0}{r}\bigg)\\ &
-8me^{4a_0lnr-4b_0lnr} \big(\frac{ a_0^2}{r^3} - \frac{ b_0}{r^2}\big)^3 + \frac{b_0 \big(1+ 4me^{2a_0lnr-2b_0lnr} \big(\frac{ a_0^2}{r^2} - \frac{ b_0}{r^2}\big)\big) }{r^2}   \bigg],
\end{split}
\end{equation}

\begin{equation}\label{19}
\begin{split}
&p_r=\frac{1}{r}e^{2a_0lnr-2b_0lnr} \bigg[  \bigg(4me^{2a_0lnr-2b_0lnr} \big(\frac{-2 a_0^2}{r^3} + \frac{2 b_0^3}{r^3}\big)+ 4me^{2a_0lnr-2b_0lnr} \big(\frac{ a_0^2}{r^2} - \frac{ b_0^2}{r^2}\big)\big(\frac{ 2 a_0}{r}- \frac{2b_0^2}{r}\big)   \bigg) \\&
\bigg(1-2a_0+2b_0  \bigg)+ \bigg( 1+ 4me^{2a_0lnr-2b_0lnr} \big(\frac{ a_0^2}{r^2} - \frac{ b_0}{r^2}\big)\big(\frac{ 4 a_0^2}{r}+ \frac{3b_0}{r}\big) \bigg) +4 r \bigg(  4me^{2a_0lnr-2b_0lnr} \big(\frac{ a_0^2}{r^2} - \frac{ b_0}{r^2}\big)  \\&
\big(\frac{ -2a_0}{r^2} + \frac{ 2b_0}{r^2}\big)+ 16 me^{2a_0lnr-2b_0lnr} \big(\frac{- a_0^2}{r^3} + \frac{ b_0}{r^3}\big) \big(\frac{ 2a_0}{r} - \frac{ 2b_0}{r}\big)
+4me^{2a_0lnr-2b_0lnr} \big(\frac{ a_0^2}{r^3} - \frac{ b_0}{r^3}\big) \big(\frac{ 2a_0}{r} - \frac{ 2b_0}{r}\big)^2\\&
24me^{2a_0lnr-2b_0lnr} \big(\frac{ a_0^2}{r^4} - \frac{ b_0}{r^4}\big)\bigg)+8rme^{4a_0lnr-4b_0lnr} \big(\frac{ a_0^2}{r^2} - \frac{ b_0}{r^2}\big)^3\bigg],
\end{split}
\end{equation}

\begin{equation}\label{20}
\begin{split}
&p_{\phi}=e^{2a_0lnr-2b_0lnr} \bigg[ 2 \bigg( 24me^{2a_0lnr-2b_0lnr} \big(\frac{ a_0^2}{r^4} - \frac{ b_0}{r^4}\big) + 8me^{2a_0lnr-2b_0lnr} \big(\frac{ a_0^2}{r^2} - \frac{ b_0}{r^2}\big) \big(\frac{- a_0}{r^2} + \frac{ b_0}{r^2}\big)\\&
+32me^{2a_0lnr-2b_0lnr} \big(\frac{ -a_0^2}{r^3} + \frac{ b_0}{r^3}\big) \big(\frac{ a_0}{r} - \frac{ b_0}{r}\big)+ 16me^{2a_0lnr-2b_0lnr} \big(\frac{ a_0^2}{r^2} - \frac{ b_0}{r^2}\big) \big(\frac{ a_0^2}{r} - \frac{ b_0}{r}\big)^2\bigg) \\&
+8me^{4a_0lnr-4b_0lnr} \big(\frac{ a_0^2}{r^2} - \frac{ b_0}{r^2}\big)^3+\bigg(1+ 4me^{2a_0lnr-2b_0lnr} \big(\frac{ a_0^2}{r^2} - \frac{ b_0}{r^2}\big)  \bigg)\bigg(\frac{- a_0}{r^2} - 4\frac{ a_0^2}{r^2}-\frac{ b_0}{r^2} + \frac{ 1}{r} \bigg)+\\&
\bigg( 8me^{2a_0lnr-2b_0lnr} \big(\frac{- a_0^2}{r^3} + \frac{ b_0}{r^2}\big)  +8me^{2a_0lnr-2b_0lnr} \big(\frac{a_0^2}{r^3} - \frac{ b_0}{r^2}\big) \big(\frac{a_0}{r} - \frac{ b_0}{r}\big) \bigg) \bigg(\frac{a_0}{r} +4 \frac{ b_0}{r}\bigg)\bigg],
\end{split}
\end{equation}

\begin{equation}\label{21}
\begin{split}
&p_{z}=e^{2a_0lnr-2b_0lnr} \bigg[ 2 \bigg( 24me^{2a_0lnr-2b_0lnr} \big(\frac{ a_0^2}{r^4} - \frac{ b_0}{r^4}\big) + 8me^{2a_0lnr-2b_0lnr} \big(\frac{ a_0^2}{r^2} - \frac{ b_0}{r^2}\big) \big(\frac{- a_0}{r^2} + \frac{ b_0}{r^2}\big)\\&
+32me^{2a_0lnr-2b_0lnr} \big(\frac{ -a_0^2}{r^3} + \frac{ b_0}{r^3}\big) \big(\frac{ a_0}{r} - \frac{ b_0}{r}\big)+ 16me^{2a_0lnr-2b_0lnr} \big(\frac{ a_0^2}{r^2} - \frac{ b_0}{r^2}\big) \big(\frac{ a_0^2}{r} - \frac{ b_0}{r}\big)^2\bigg) \\&
+ \bigg( 8me^{2a_0lnr-2b_0lnr} \big(\frac{ -a_0^2}{r^3} + \frac{ b_0}{r^2}\big)+8me^{2a_0lnr-2b_0lnr} \big(\frac{ a_0^2}{r^3} - \frac{ b_0}{r^2}\big)  \big(\frac{ 2a_0}{r} - \frac{ 2b_0}{r}\big)   \bigg)\\&
+8me^{4a_0lnr-4b_0lnr} \big(\frac{ a_0^2}{r^2} - \frac{ b_0}{r^2}\big)^3-\frac{b_0\bigg(1+4me^{4a_0lnr-4b_0lnr} \big(\frac{ a_0^2}{r^2} - \frac{ b_0}{r^2}\big)  \bigg)  }{r^2}\bigg],
\end{split}
\end{equation}

\begin{figure}[h!]
\centering
\epsfig{file=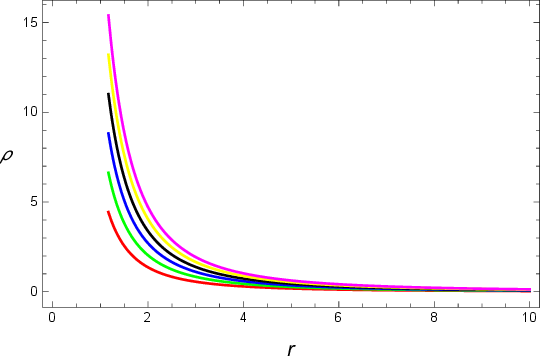, width=.32\linewidth,height=2in}
~~~~~~~~~~\epsfig{file=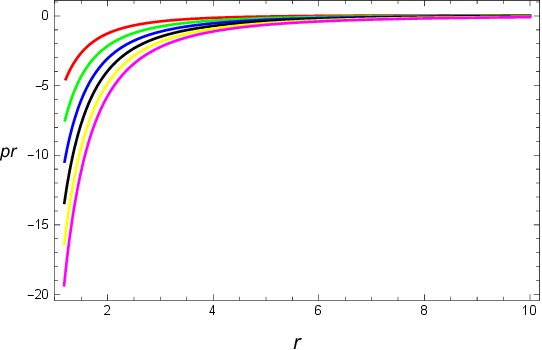, width=.32\linewidth,height=2in}\\
\epsfig{file=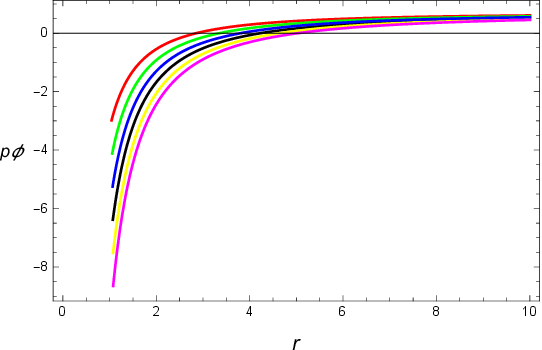, width=.32\linewidth,height=2in}
~~~~~~~~~~\epsfig{file=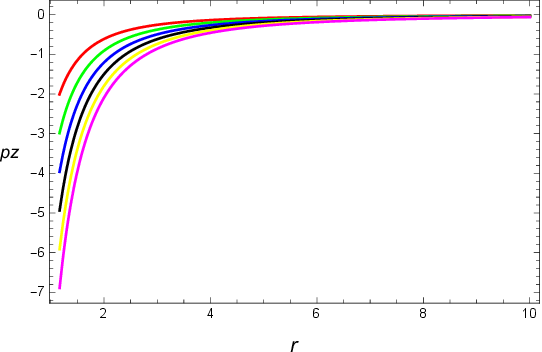, width=.32\linewidth,height=2in}
\caption{Graphical representation of energy density and pressure components of Levi Civita Solution.}
\label{Fig.1}
\end{figure}

\begin{figure}[h!]
\centering
\epsfig{file=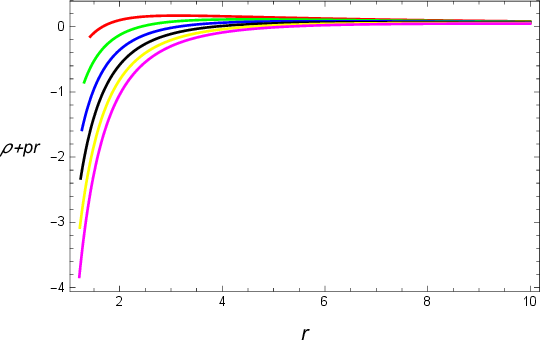, width=.32\linewidth,height=2in}
~~\epsfig{file=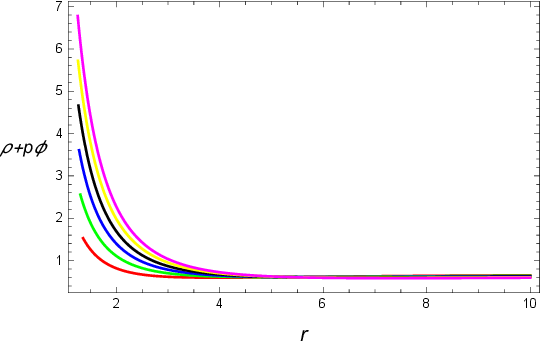, width=.32\linewidth,height=2in}
~~\epsfig{file=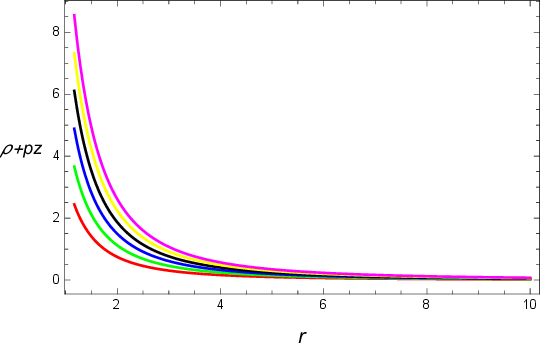, width=.32\linewidth,height=2in}
\caption{Graphical representation of NEC components of Levi Civita Solution.}
\label{Fig.2}
\end{figure}

\begin{figure}[h!]
\centering
\epsfig{file=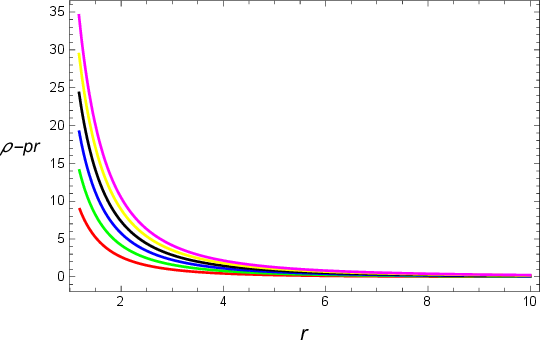, width=.32\linewidth,height=2in}
~~\epsfig{file=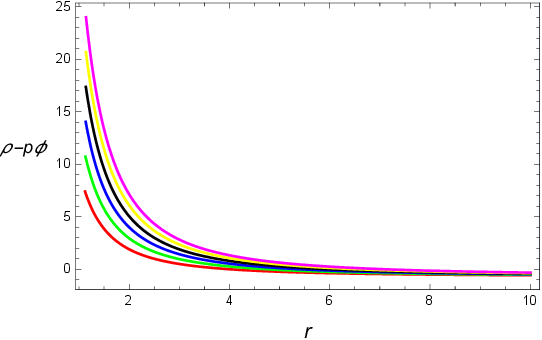, width=.32\linewidth,height=2in}
~~\epsfig{file=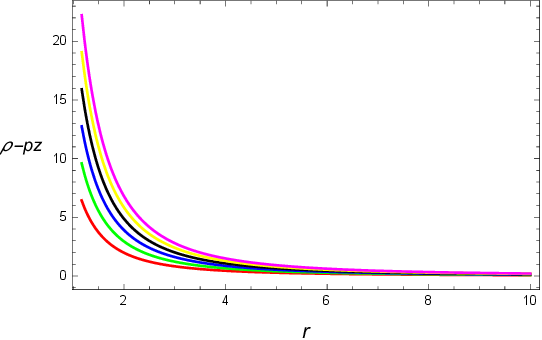, width=.32\linewidth,height=2in}
\caption{Graphical representation of DEC components of Levi Civita Solution.}
\label{Fig.3}
\end{figure}

\begin{figure}[h!]
\centering
\epsfig{file=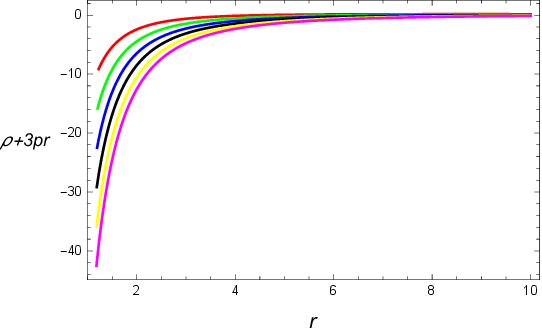, width=.32\linewidth,height=2in}
~~\epsfig{file=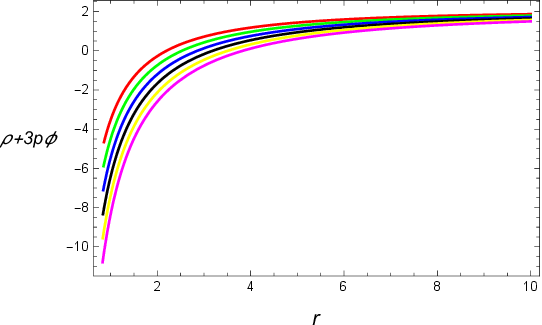, width=.32\linewidth,height=2in}
~~\epsfig{file=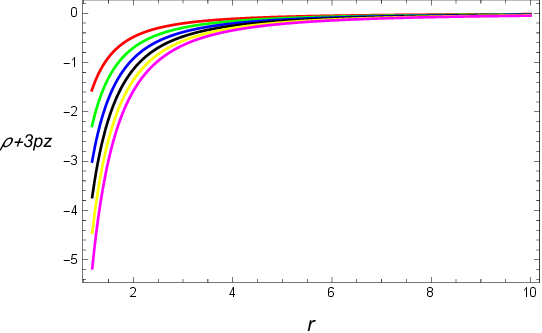, width=.32\linewidth,height=2in}
\caption{Graphical representation of SEC components of Levi Civita Solution.}
\label{Fig.4}
\end{figure}
\subsection{Discussion}
In this subsection, we observe the graphical evaluation of energy density, pressure components and energy conditions of Levi civita spacetime. The graphical representation of energy density is positive and decreasing in nature, when we moves towards the radial coordinate as shown in Figure (\ref{Fig.1}). The other three pressure components including radial pressure $p_r$, azimuthal pressure $p_{\phi}$, and axial pressure $p_z$ have negative trends and increasing in nature as shown in Figure (\ref{Fig.1}). Figure (\ref{Fig.2}) shows that the graphical analysis of the sum of energy density and radial pressure is negative and increasing in nature, while the other two components i.e., sum of energy density with azimuthal pressure and sum of energy density with axial pressure are positive and approaches to zero as seen in Figure (\ref{Fig.2}). It can be concluded that NEC is violated due to the negative nature of $\rho+p_r$. As, we know that WEC is the the combination of energy density and the components of NEC, so, one can easily summarized that WEC is also not satisfied. The graphical analysis of three components of DEC including the difference of energy density with radial pressure, difference of energy density with azimuthal pressure and difference of energy density with axial pressure are positive and decreasing, which means DEC is satisfied for this case. Moreover, the graphical representation of all the components of SEC i.e., the sum of energy density with three times of radial pressure, the sum of energy density with three times of azimuthal pressure, and the sum of energy density with three times of axial pressure are negative, which means that SEC is violated for Levia civita case. Upon examination, the results lead to the observation that null energy conditions are breached within distinct regions. This observation serves as a noteworthy indication, strongly implying the possible existence of cylindrical wormholes in those specific areas. The implications of these findings open avenues for further exploration and understanding within the realm of theoretical physics.

%

\section{Cosmic String Case}
 Cylindrically symmetric solutions play a crucial role in understanding the significance of cosmic strings, hypothetical one-dimensional topological defects that may have formed during the early universe's phase transitions. These solutions are essential for analyzing the gravitational effects of cosmic strings on surrounding matter and energy distributions, contributing to predictions related to the cosmic microwave background, large-scale structure, and gravitational lensing. Furthermore, the study of cylindrically symmetric solutions provides insights into the topology and geometry of spacetime around cosmic strings, helping to decipher their interactions and behaviors over cosmic time scales. These solutions also form a fundamental part of the theoretical framework for modeling the formation and evolution of cosmic string networks, including the dynamics of loop formation and gravitational wave emission. Notably, the presence of cosmic strings with cylindrical symmetry leaves distinct imprints on the cosmic microwave background, and understanding these imprints is crucial for interpreting observational data and placing constraints on the properties of cosmic strings, contributing to advancements in both theoretical cosmology and observational astrophysics. We have discussed this case under the application of $f(R)$ theory of gravity for the following metric potentials
 \begin{equation}\label{22}
a(r)=A_{0}ln[w(r)],~~~b(r)=b_{0}ln[w(r)],~~~w(r)=c_{0}r,
 \end{equation}
 where, $a_{0}, b_{0}$ and $c_{0}$ are any arbitrary constants. Also, if we put $c_{0}=1$ in Eq. (\ref{22}) then it becomes LC solutions defined in (\ref{57}). By taking this, we obtain
\begin{equation}\label{23}
dS^2=(c_0r)^{2a_0}dt^2-(c_0r)^{2-2a_0}d\phi^2-(c_0r)^{2b_0-2a_0}(dr^2+dz^2).
\end{equation}
 Now, putting these metric potentials in Eqs. (\ref{14})-(\ref{17}), we have the following equations as

\begin{equation}\label{24}
\begin{split}
&\rho=e^{2a_0lnc_0r-2b_0lnc_0r}\bigg[-2 \bigg( 24m e^{2a_0lnc_0r-2b_0lnc_0r} \big(\frac{ a_0^2}{r^4} - \frac{ b_0}{r^4}\big) -8m e^{2a_0lnc_0r-2b_0lnc_0r} \big(\frac{ a_0^2}{r^2} - \frac{ b_0}{r^2}\big)\big(\frac{ -a_0 }{r^2} +\frac{ b_0}{r^2}\big)\\&
+16m e^{2a_0lnc_0r-2b_0lnc_0r} \big(\frac{ -a_0^2}{r^3} + \frac{ b_0}{r^3}\big)\big(\frac{ a_0 }{r} -\frac{ b_0}{r}\big)+ 16m e^{2a_0lnc_0r-2b_0lnc_0r} \big(\frac{ a_0^2}{r^2}- \frac{ b_0}{r^2}\big)\big(\frac{ a_0 }{r} -\frac{ b_0}{r}\big)^2\bigg)\\&
+ \bigg( 8m e^{2a_0lnc_0r-2b_0lnc_0r} \big(\frac{ -a_0^2}{r^3} + \frac{ b_0}{r^3}\big) + 8m e^{2a_0lnc_0r-2b_0lnc_0r} \big(\frac{ a_0^2}{r^2} - \frac{ b_0}{r^2}\big)\big(\frac{ a_0}{r} - \frac{ b_0}{r}\big) \bigg)\bigg(\frac{ a_0}{r} - \frac{ b_0}{r}\bigg)\\&
-8m e^{2a_0lnc_0r-2b_0lnc_0r} \big(\frac{ a_0^2}{r^2} - \frac{ b_0}{r^2}\big)^3 +\frac{b_0\big(1+4m e^{2a_0lnc_0r-2b_0lnc_0r} \big(\frac{ a_0^2}{r^2} - \frac{ b_0}{r^2}\big) \big) }{r^2}  \bigg],
\end{split}
\end{equation}

\begin{equation}\label{25}
\begin{split}
&p_r=\frac{1}{c_0 r}e^{2a_0lnc_0r-2b_0lnc_0r}\bigg[\bigg(c_0 + 2 (-a_0 c_0 + b_0 c_0)\bigg)\bigg( 8m e^{2a_0lnc_0r-2b_0lnc_0r} \big(\frac{ -a_0^2}{r^3} + \frac{ b_0}{r^3}\big)\\&
+8m e^{2a_0lnc_0r-2b_0lnc_0r} \big(\frac{ a_0^2}{r^2}- \frac{ b_0}{r^2}\big)\big(\frac{ a_0^2}{r}- \frac{ b_0}{r}\big)\bigg)+ c_0 \bigg(1+m e^{2a_0lnc_0r-2b_0lnc_0r} \big(\frac{ a_0^2}{r^2}- \frac{ b_0}{r^2}\big)  \bigg) \bigg(\frac{ 4a_0^2}{r}+ \frac{ 3 b_0}{r}\bigg)\\&
+4 c_0 \bigg( 24m e^{2a_0lnc_0r-2b_0lnc_0r} \big(\frac{ a_0^2}{r^4}- \frac{ b_0}{r^4}\big) +8m e^{2a_0lnc_0r-2b_0lnc_0r} \big(\frac{ a_0^2}{r^2}- \frac{ b_0}{r^2}\big)\big(\frac{ -a_0}{r^2}+ \frac{ b_0}{r^2}\big) \\&
+16m e^{2a_0lnc_0r-2b_0lnc_0r} \big(\frac{ -a_0^2}{r^3}+ \frac{ b_0}{r^3}\big)\big(\frac{ a_0}{r^2}- \frac{ b_0}{r^2}\big)+16rm e^{2a_0lnc_0r-2b_0lnc_0r} \big(\frac{ a_0^2}{r^2}- \frac{ b_0}{r^2}\big)\big(\frac{ a_0}{r^2}- \frac{ b_0}{r^2}\big)^2 \bigg)\\&
+8rm e^{2a_0lnc_0r-2b_0lnc_0r} \big(\frac{ a_0^2}{r^2}- \frac{ b_0}{r^2}\big)^3 \bigg],
\end{split}
\end{equation}

\begin{equation}\label{26}
\begin{split}
&p_\phi=e^{2a_0lnc_0r-2b_0lnc_0r}\bigg[2\bigg( 24m e^{2a_0lnc_0r-2b_0lnc_0r} \big(\frac{ a_0^2}{r^4} - \frac{ b_0}{r^4}\big)
+8m e^{2a_0lnc_0r-2b_0lnc_0r} \big(\frac{ a_0^2}{r^2}- \frac{ b_0}{r^2}\big)\big(\frac{ -a_0}{r^2}+ \frac{ b_0}{r^2}\big)\\&
+32m e^{2a_0lnc_0r-2b_0lnc_0r} \big(\frac{- a_0^2}{r^3}+ \frac{ b_0}{r^3}\big)\big(\frac{ a_0}{r}- \frac{ b_0}{r}\big)
+16m e^{2a_0lnc_0r-2b_0lnc_0r} \big(\frac{ a_0^2}{r^2}- \frac{ b_0}{r^2}\big)\big(\frac{ a_0}{r}- \frac{ b_0}{r}\big)^2\bigg)\\&
+8m e^{2a_0lnc_0r-2b_0lnc_0r} \big(\frac{ a_0^2}{r^2}- \frac{ b_0}{r^2}\big)^3+
\bigg(1+4m e^{2a_0lnc_0r-2b_0lnc_0r} \big(\frac{ a_0^2}{r^2}- \frac{ b_0}{r^2}\big) \bigg)\bigg(\frac{ -a_0}{r^2}-\frac{ 4a_0^2}{r^2}- \frac{ b_0}{r^2}+\frac{ 1}{r}\bigg)\\&
+ \bigg(4m e^{2a_0lnc_0r-2b_0lnc_0r} \big(\frac{ -2a_0^2}{r^3}+ \frac{ b_0}{r^3}\big)
+8m e^{2a_0lnc_0r-2b_0lnc_0r} \big(\frac{ a_0^2}{r^2}- \frac{ b_0}{r^2}\big)\big(\frac{ a_0}{r}- \frac{ b_0}{r}\big) \bigg)\bigg(\frac{ a_0}{r}+4 \frac{ b_0}{r}\bigg) \bigg],
\end{split}
\end{equation}

\begin{equation}\label{27}
\begin{split}
&p_z=e^{2a_0lnc_0r-2b_0lnc_0r}\bigg[2\bigg( 24m e^{2a_0lnc_0r-2b_0lnc_0r} \big(\frac{ a_0^2}{r^4} - \frac{ b_0}{r^4}\big)
+8m e^{2a_0lnc_0r-2b_0lnc_0r} \big(\frac{ a_0^2}{r^2}- \frac{ b_0}{r^2}\big)\big(\frac{ -a_0}{r^2}+ \frac{ b_0}{r^2}\big)\\&
+32m e^{2a_0lnc_0r-2b_0lnc_0r} \big(\frac{- a_0^2}{r^3}+ \frac{ b_0}{r^3}\big)\big(\frac{ a_0}{r}- \frac{ b_0}{r}\big)
+16m e^{2a_0lnc_0r-2b_0lnc_0r} \big(\frac{ a_0^2}{r^2}- \frac{ b_0}{r^2}\big)\big(\frac{ a_0}{r}- \frac{ b_0}{r}\big)^2\bigg)\\&
+2 \bigg(8m e^{2a_0lnc_0r-2b_0lnc_0r} \big(\frac{ -a_0^2}{r^3}+ \frac{ b_0}{r^3}\big)
+8m e^{2a_0lnc_0r-2b_0lnc_0r} \big(\frac{ -a_0^2}{r^3}+ \frac{ b_0}{r^3}\big) \big(\frac{ 2a_0}{r}- \frac{ b_0}{r}\big) \bigg)\bigg(\frac{1}{r}-\frac{ a_0}{r}+ \frac{ b_0}{r} \bigg) \\&
+8m e^{2a_0lnc_0r-2b_0lnc_0r} \big(\frac{ a_0^2}{r^2}- \frac{ b_0}{r^2}\big)^3
- \frac{b_0 \big( 1+4 m e^{2a_0lnc_0r-2b_0lnc_0r} \big(\frac{ a_0^2}{r^2} - \frac{ b_0}{r^2}\big) \big)}{r^2}\bigg],
\end{split}
\end{equation}
 These set of  Eqs. (\ref{24})-(\ref{27}) involves some constants $a_{0}, b_{0}$ and $c_{0}$. If we restrict these constants to some values like $a_{0}=0, b_{0}=0$ and $c_{0}$ only lies between the interval $(0, 1)$ then the space-time is corresponding to the exterior metric of a cosmic string with the following line element
\begin{equation}\label{28}
ds^2=dt^2-(c_0r)^2d\phi^2-dr^2-dz^2.
\end{equation}

The mathematical expression provided in Equation (\ref{28}) describes the metric that reveals the configuration surrounding a linear cosmic string, reminiscent of flat space-time. Through the exploration of Ricci flat solutions, we derived a conical space-time, specifically a zero curvature space-time, as a particular instance. This encompasses the space-time associated with a cosmic string. Cosmic strings, theoretical one-dimensional structures embedded in the fabric of spacetime, are postulated entities arising from certain cosmological models, particularly those involving phase transitions in the early universe. When examining cosmic string solutions with a focus on cylindrical symmetry, researchers delve into vortex-like configurations, where the properties of these strings maintain consistency along their length, akin to a straight wire. This cylindrically symmetric approach simplifies the mathematical formulations, facilitating a more accessible analysis of cosmic string behavior. When discussing these solutions, it is imperative to uphold academic integrity, avoiding plagiarism and providing proper attribution to the original contributors in the field. By approaching the exploration of cosmic string solutions with ethical rigor, researchers contribute to the advancement of our understanding of the universe while respecting the intellectual contributions of their peers in the scientific community.

%
\begin{figure}[h!]
\centering
\epsfig{file=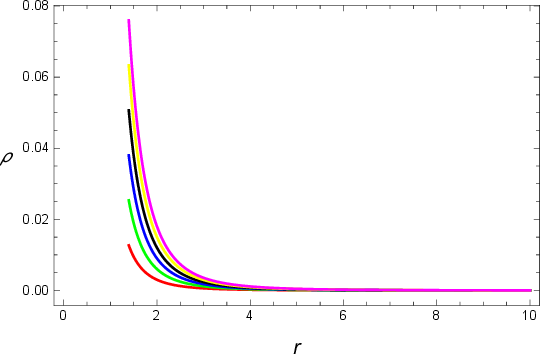, width=.32\linewidth,height=2in}
~~~~~~~~~~\epsfig{file=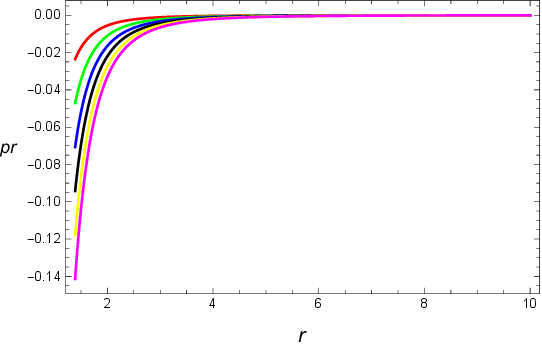, width=.32\linewidth,height=2in}\\
\epsfig{file=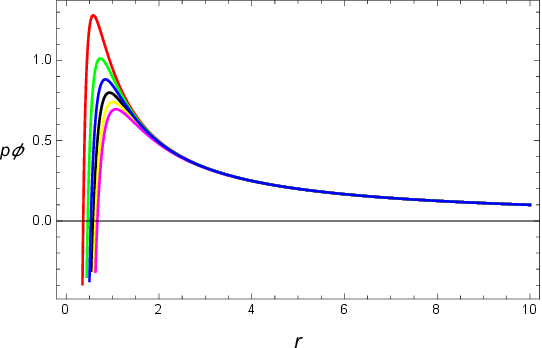, width=.32\linewidth,height=2in}
~~~~~~~~~~\epsfig{file=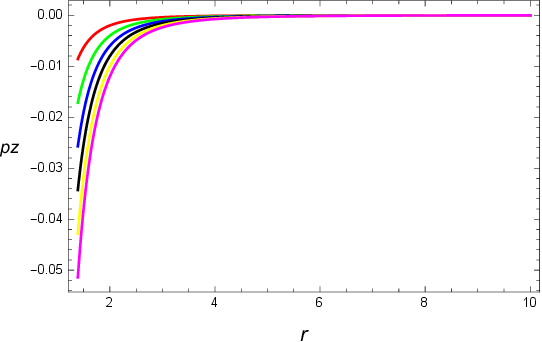, width=.32\linewidth,height=2in}
\caption{Graphical representation of energy density and pressure components of Cosmic String Solution.}
\label{Fig.5}
\end{figure}

\begin{figure}[h!]
\centering
\epsfig{file=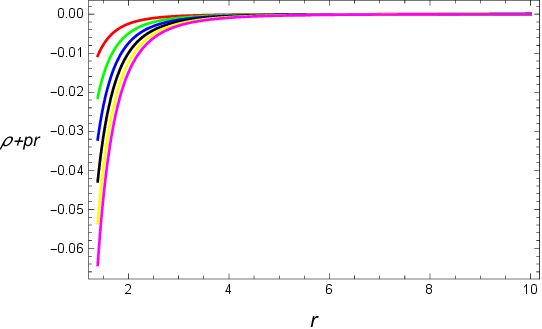, width=.32\linewidth,height=2in}
~~\epsfig{file=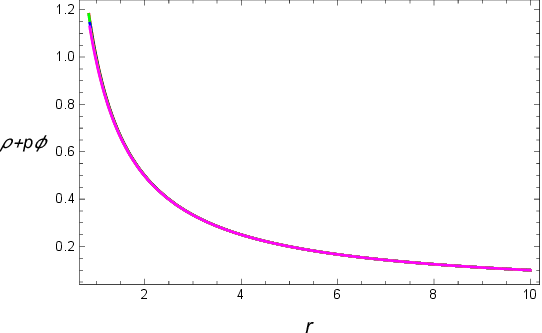, width=.32\linewidth,height=2in}
~~\epsfig{file=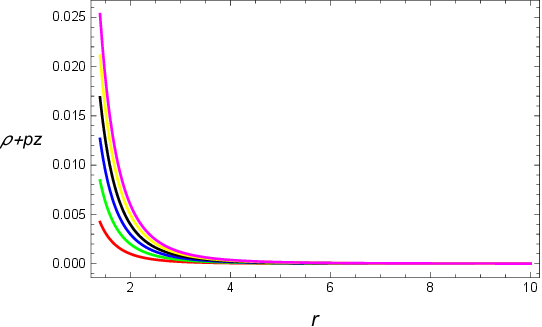, width=.32\linewidth,height=2in}
\caption{Graphical representation of NEC components of Cosmic String Solution.}
\label{Fig.6}
\end{figure}

\begin{figure}[h!]
\centering
\epsfig{file=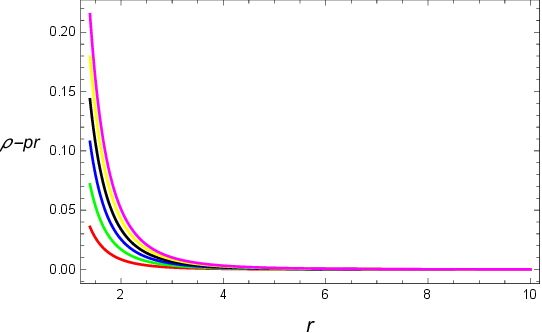, width=.32\linewidth,height=2in}
~~\epsfig{file=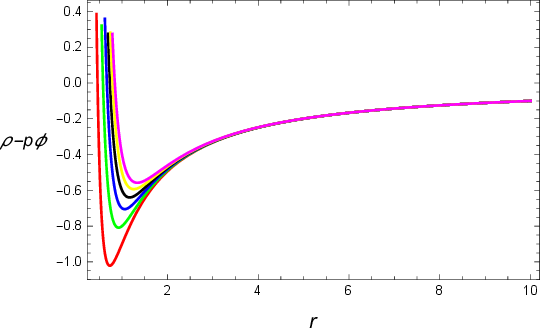, width=.32\linewidth,height=2in}
~~\epsfig{file=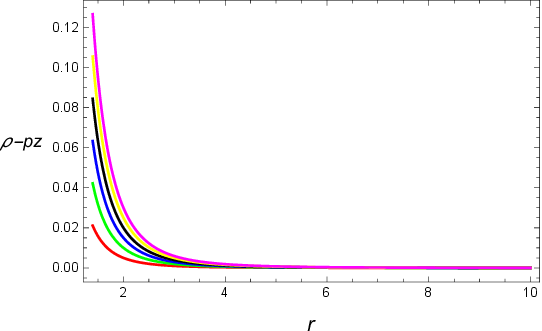, width=.32\linewidth,height=2in}
\caption{Graphical representation of DEC components of Cosmic String Solution.}
\label{Fig.7}
\end{figure}

\begin{figure}[h!]
\centering
\epsfig{file=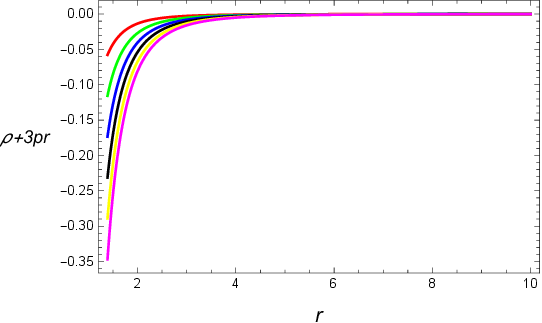, width=.32\linewidth,height=2in}
~~\epsfig{file=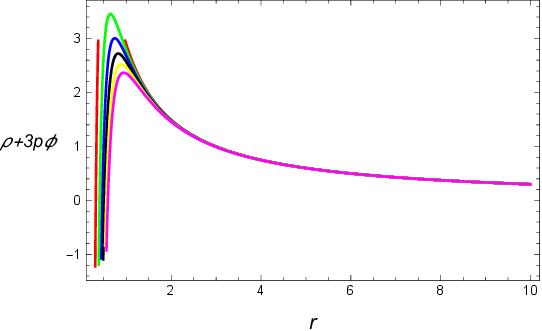, width=.32\linewidth,height=2in}
~~\epsfig{file=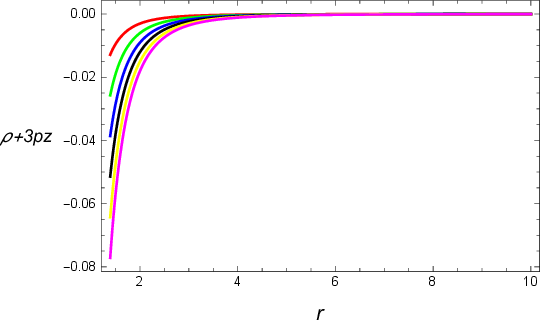, width=.32\linewidth,height=2in}
\caption{Graphical representation of SEC components of Cosmic String Solution.}
\label{Fig.8}
\end{figure}

\subsection{Discussion}

This subsection will provide the detailed graphical analysis of various aspects, including energy density, pressure components, and energy conditions within the cosmic string spacetime. The graphical depiction of energy density exhibits a positive and decreasing trend as we progress along the radial coordinate, as illustrated in Figure (\ref{Fig.5}). Conversely, the radial pressure ($p_r$), and and axial pressure ($p_z$) show negative trends, with an increasing nature, as depicted in the same figure. Moreover, the graphical analysis of azimuthal pressure ($p_{\phi}$) is initially negative and increasing but it shows negative behavior when we moves towards the boundary as seen in Figure (\ref{Fig.5}). Figure (\ref{Fig.6}) presents an analysis where the sum of energy density and radial pressure is portrayed as negative and increasing, while the sums involving azimuthal pressure and axial pressure are positive, converging towards zero. This observation leads to the conclusion that the null energy condition (NEC) is violated due to the negativity of $\rho+p_r$. Given that the Weak Energy Condition (WEC) relies on the combination of energy density and NEC components, it follows that WEC is not satisfied in this scenario. Further scrutiny of the graphical analysis for the components of the dominant energy condition (DEC) reveals positive and decreasing trends for the differences involving energy density with radial pressure, and axial pressure. This suggests that DEC is indeed satisfied in this particular case as shown in Figure (\ref{Fig.7}). Moreover, the graphical analysis of the difference of energy density with azimuthal pressure is initially decreasing and negative but it shows increasing nature when $r>1$ as seen in the Figure (\ref{Fig.7}), which means that DEC is violated for cosmic string case. Moving on to the strong energy condition (SEC), the graphical representation demonstrates negativity in all components involving the sum of energy density with three times radial pressure, azimuthal pressure, and axial pressure. This implies a violation of SEC for the cosmic string case. In summary, the graphical evaluations indicate a breach of null energy conditions in specific regions, strongly hinting at the potential existence of cylindrical wormholes in those areas. These findings hold significant implications for further exploration and understanding within the field of theoretical physics.


\section{Cylindrically Symmetric Solutions}
In this section, we are going to discuss few more cylindrically symmetric solutions in modified $f(R)$ theory of gravity. Cylindrically symmetric solutions in $f(R)$ theory of gravity are significant for addressing the cosmos because they offer a framework for studying gravitational interactions in non-standard theories of gravity beyond Einstein's general relativity. These solutions are particularly relevant in contexts where cylindrical symmetry is present, such as in the study of galaxies, cosmic filaments, and other large-scale cosmic structures. By examining how $f(R)$ theory of gravity gravity behaves in these situations, researchers can gain insights into the potential modifications of gravitational dynamics on cosmic scales. This understanding is crucial for accurately modeling the behavior of the universe, especially in regions where traditional general relativity might not provide a complete description. Furthermore, by studying cylindrically symmetric solutions, scientists can assess the compatibility of $f(R)$ theory of gravity, allowing for a more comprehensive and nuanced understanding of the underlying gravitational physics governing the cosmos. A few more cases are discussed in coming subsections are as follows:

\subsection{Case-I}

In this case, we have done our calculations by using the following metric potentials
\begin{equation}\label{29}
a(r)=a_{0}r+ln[w(r)],~~~b(r)=ln[w(r)],~~~w(r)=c_{0}r,
 \end{equation}
 where, $a_{0}, b_{0}$ and $c_{0}$ are any arbitrary constants. Substituting these metric potentials in Eqs. (\ref{14})-(\ref{17}), we get

\begin{equation}\label{30}
\begin{split}
&\rho=\frac{1}{(w_0+r w_1)^{12}}\bigg[ w_1^2 (w_0 + r w_1)^{10} - 72 a_0^5 m w_0^3 w_1 (w_0 + r w_1) (2 w_0 + 3 r w_1)^2-8 a_0^6 m w_0^3 (2 w_0 + 3 r w_1)^3
\\ & +a_0 w_1 (w_0 + r w_1)^7 (-3 w_0^3 - 4 r w_0^2 w_1 + (-348 m + r^2) w_0 w_1^2 + 2 r^3 w_1^3) - a_0^2 (w_0 + r w_1)^6 (3 w_0^4 + 9 r w_0^3 w_1
\\& +4 (103 m + 2 r^2) w_0^2 w_1^2 + r (1044 m + r^2) w_0 w_1^3 - r^4 w_1^4) - 4 (a_0^3) m w_0 w_1 ((w_0 + r w_1)^3)(45 w_0^4 + 296 r w_0^3 w_1
\\& +2 (27 + 359 r^2) w_0^2 w_1^2 + 728 r^3 w_0 w_1^3 + 261 r^4 w_1^4) - 4 a_0^4 m w_0 (w_0 + r w_1)^2 (6 w_0^5 + 57 r w_0^4 w_1 + (108 +
\\& 199 r^2) w_0^3 w_1^2 +2 r (81 + 169 r^2) w_0^2 w_1^3 + 277 r^4 w_0 w_1^4 + 87 r^5 w_1^5)) \bigg].
 \end{split}
   \end{equation}

   \begin{equation}\label{31}
\begin{split}
&p_r=\frac{1}{(w_0+r w_1)^{12}}\bigg[ 7 w_1^2 (w_0 + r w_1)^10 + 72 a_0^5 m w_0^3 w_1 (w_0 + r w_1) (2 w_0 + 3 r w_1)^2 + 8 a_0^6 m w_0^3 (2 w_0 + 3 r w_1)^3
\\ & +a_0 w_1 (w_0 + r w_1)^7 (7 w_0^3 +  28 r w_0^2 w_1 + (624 m + 35 r^2) w_0 w_1^2 + 14 r^3 w_1^3) + ( a_0^2) ((w_0 + r w_1)^6)(3 w_0^4 + 13 r w_0^3 w_1
\\&  +8 (56 m + 3 r^2) w_0^2 w_1^2 + 3 r (624 m + 7 r^2) w_0 w_1^3 +    7 r^4 w_1^4) + 4 a_0^3 m w_0 w_1 (w_0 + r w_1)^3 (33 w_0^4 + 290 r w_0^3 w_1 +(54
\\& +949 r^2) w_0^2 w_1^2 + 1160 r^3 w_0 w_1^3 + 468 r^4 w_1^4) + 4 a_0^4 m w_0 (w_0 + r w_1)^2 (6 w_0^5 + 45 r w_0^4 w_1 + 4 (27 +46 r^2) w_0^3 w_1^2
\\& +r (162 + 413 r^2) w_0^2 w_1^3 + 424 r^4 w_0 w_1^4 + 156 r^5 w_1^5)  \bigg].
\end{split}
   \end{equation}

   \begin{equation}\label{32}
\begin{split}
& p_\phi=\frac{1}{(w_0+r w_1)^{12}}\bigg[w_1 (w_0 + (-6 + r) w_1) (w_0 + r w_1)^10 + 72 a_0^5 m w_0^3 w_1 (w_0 + r w_1) (2 w_0 + 3 r w_1)^2 + 8 a_0^6 m w_0^3 (
\\& 2 w_0 + 3 r w_1)^3+2 a_0 w_1 (w_0 + r w_1)^7 ((-5 + r) w_0^3 + (6 m + r (-16 + 3 r)) w_0^2 w_1 + (6 m (3 + r) +r^2 (-17 + 3 r)) w_0 w_1^2
\\& +(-6 + r) r^3 w_1^3) - a_0^2 (w_0 + r w_1)^6 (5 w_0^4 - (8 m + (-20 + r) r) w_0^3 w_1 + (-44 m (-4 + r) + (31 - 3 r) r^2) w_0^2 w_1^2
\\& -r (36 m (3 + r) + r^2 (-22 + 3 r)) w_0 w_1^3 - (-6 + r) r^4 w_1^4) + 4 a_0^3 m w_0 w_1 (w_0 + r w_1)^3 (4 (-10 + r) w_0^4 + 21 (-8
\\& + r) r w_0^3 w_13 (18 + r^2 (-63 + 13 r)) w_0^2 w_1^2 + r^3 (-34 + 31 r) w_0 w_1^3 + 9 r^4 (3 + r) w_1^4) - 4 a_0^4 m w_0 (w_0 + r w_1)^2 (10 w_0^5
\\& -2 (-30 + r) r w_0^4 w_1 + (-108 + (134 - 9 r) r^2) w_0^3 w_1^2 + r (-162 + (119 - 15 r) r^2) w_0^2 w_1^3 + (26 - 11 r) r^4 w_0 w_1^4
\\&- 3 r^5 (3 + r) w_1^5)
    \bigg].
\end{split}
   \end{equation}

   \begin{equation}\label{33}
\begin{split}
& p_z=-\frac{1}{(w_0+r w_1)^{12}}\bigg[w_1^2 (w_0 + r w_1)^10 - 72 a_0^5 m w_0^3 w_1 (w_0 + r w_1) (2 w_0 + 3 r w_1)^2 - 8 a_0^6 m w_0^3 (2 w_0 + 3 r w_1)^3
\\& +a_0 w_1 (w_0 + r w_1)^7 (w_0^3 + 4 r w_0^2 w_1 + (-204 m + 5 r^2) w_0 w_1^2 +  2 r^3 w_1^3) + a_0^2 (w_0 + r w_1)^6 (w_0^4 + 3 r w_0^3 w_1
\\& +4 (-35 m + r^2) w_0^2 w_1^2 + 3 r (-204 m + r^2) w_0 w_1^3 + r^4 w_1^4) - 4 a_0^3 m w_0 w_1 (w_0 + r w_1)^3 (5 w_0^4+80 r w_0^3 w_1
\\& +2 (27 + 149 r^2) w_0^2 w_1^2 + 376 r^3 w_0 w_1^3 + 153 r^4 w_1^4) + 4 (a_0^4) m w_0 (w_0 + r w_1)^2 (2 w_0^5 - r w_0^4 w_1-(108
\\& +43 r^2) w_0^3 w_1^2 - 18 r (9 + 7 r^2) w_0^2 w_1^3 - 137 r^4 w_0 w_1^4 -  51 r^5 w_1^5)\bigg].
\end{split}
   \end{equation}

\begin{figure}[h!]
\centering
\epsfig{file=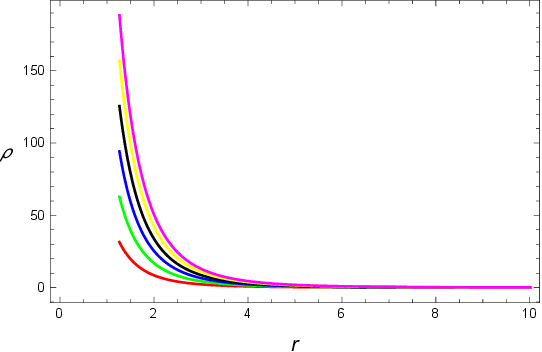, width=.32\linewidth,height=2in}
~~~~~~~~~~\epsfig{file=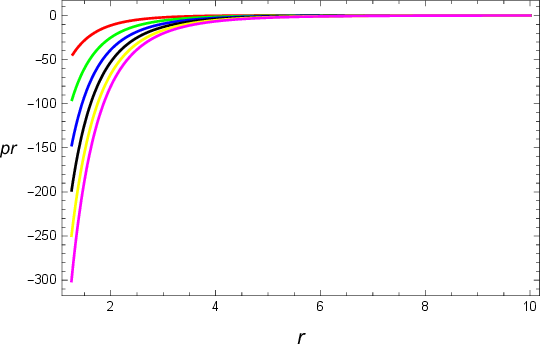, width=.32\linewidth,height=2in}\\
\epsfig{file=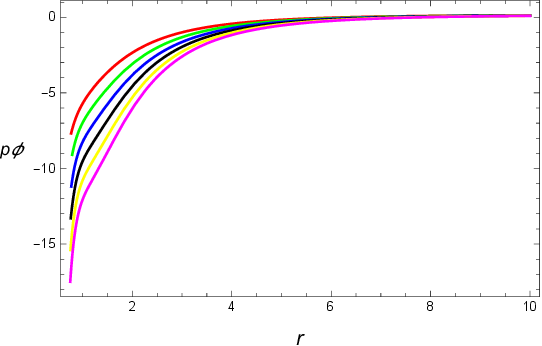, width=.32\linewidth,height=2in}
~~~~~~~~~~\epsfig{file=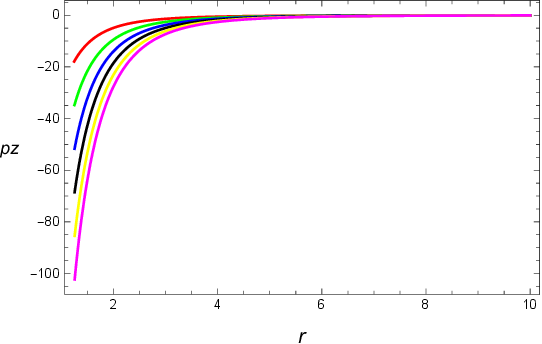, width=.32\linewidth,height=2in}
\caption{Graphical representation of energy density and pressure components for Case-I.}
\label{Fig.9}
\end{figure}

\begin{figure}[h!]
\centering
\epsfig{file=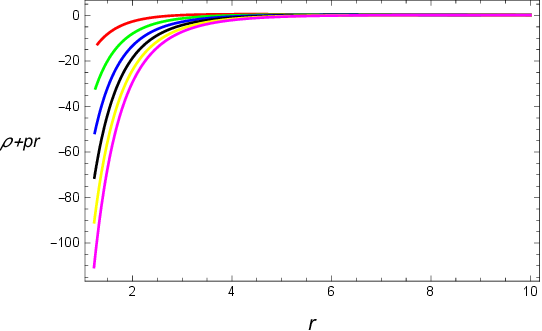, width=.32\linewidth,height=2in}
~~\epsfig{file=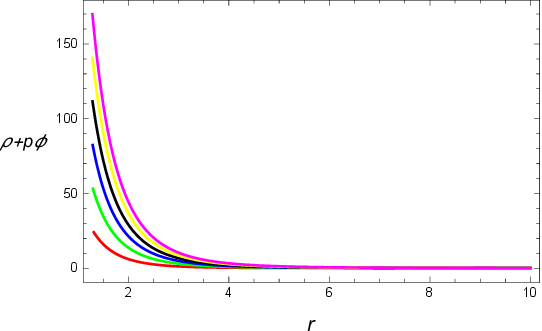, width=.32\linewidth,height=2in}
~~\epsfig{file=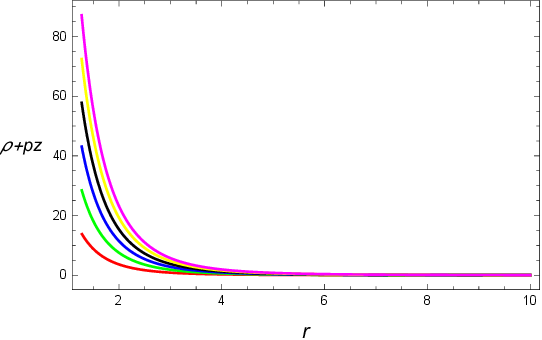, width=.32\linewidth,height=2in}
\caption{Graphical representation of NEC components of for Case-I.}
\label{Fig.10}
\end{figure}

\begin{figure}[h!]
\centering
\epsfig{file=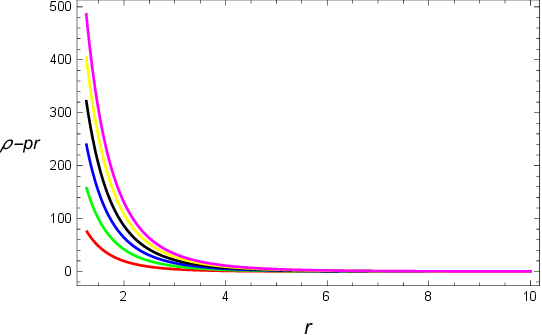, width=.32\linewidth,height=2in}
~~\epsfig{file=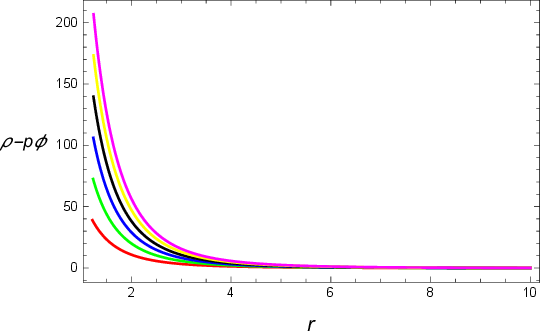, width=.32\linewidth,height=2in}
~~\epsfig{file=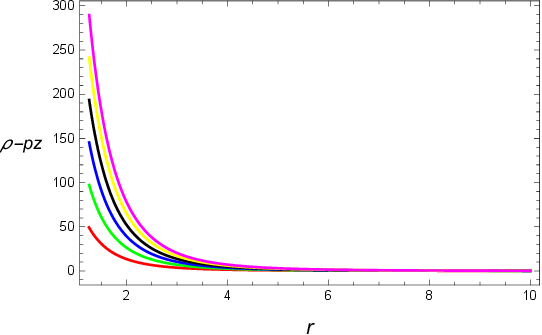, width=.32\linewidth,height=2in}
\caption{Graphical representation of DEC components for Case-I.}
\label{Fig.11}
\end{figure}

\begin{figure}[h!]
\centering
\epsfig{file=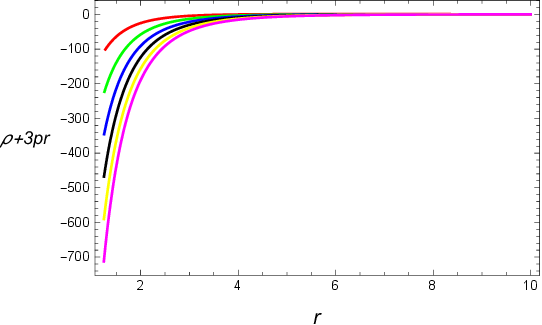, width=.32\linewidth,height=2in}
~~\epsfig{file=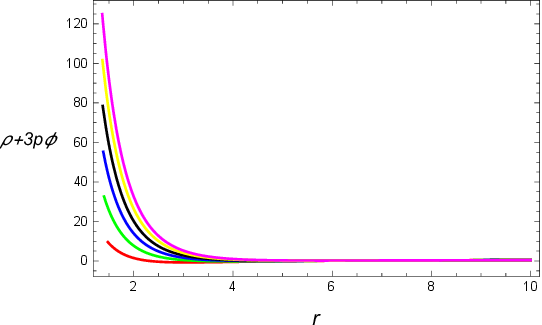, width=.32\linewidth,height=2in}
~~\epsfig{file=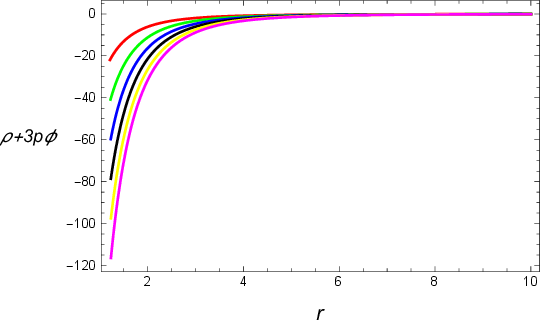, width=.32\linewidth,height=2in}
\caption{Graphical representation of DEC components for Case-I.}
\label{Fig.12}
\end{figure}

This subsection provides an in-depth graphical analysis of various elements for case-I, encompassing energy density, pressure components, and energy conditions. The graphical representation of energy density reveals a positive and decreasing trend along the radial coordinate, as illustrated in Figure (\ref{Fig.9}). Conversely, radial pressure ($p_r$), azimuthal pressure ($p_{\phi}$) and axial pressure ($p_z$) exhibit negative trends with an increasing nature in the same figure. Figure (\ref{Fig.10}) further analyzes the sum of energy density and radial pressure, portraying a negative and increasing pattern, while the sums involving azimuthal pressure and axial pressure are positive, converging towards zero. This observation leads to the inference that the null energy condition (NEC) is violated due to the negativity of $\rho+p_r$. Consequently, the Weak Energy Condition (WEC), dependent on the combination of energy density and NEC components, is not satisfied in this scenario. A closer examination of the graphical analysis for components of the dominant energy condition (DEC) reveals positive and decreasing trends for the differences involving energy density with radial pressure, azimuthal pressure and axial pressure as observed in Figure (\ref{Fig.11}), which indicates that DEC is satisfied for this case. Turning attention to the strong energy condition (SEC), the graphical representation illustrates negativity in two components related to the sum of energy density with three times radial pressure, and axial pressure. But, the graphical analysis of the sum of energy density with three times azimuthal pressure is positive as shown in Figure (\ref{Fig.10}). One can easily summarize that SEC is violated for this case. In summary, the graphical assessments suggest a breach of null energy conditions in specific regions, strongly suggesting the potential existence of cylindrical wormholes in those areas. These findings have profound implications for further exploration and understanding within the realm of theoretical physics.

\subsection{Case-II}

\begin{equation}\label{34}
a(r)=ln[w(r)+a_{0} R^2],~~~b(r)=ln[w(r)],~~~w(r)=w_{0},
 \end{equation}
 where, $a_{0}, b_{0}$ and $c_{0}$ are any arbitrary constants. Substituting these metric potentials in Eqs. (\ref{14})-(\ref{17}), we get

\begin{equation}\label{35}
\begin{split}
&\rho=\frac{1}{w_0^6} 2 a_0 \bigg[-864 a_0^5 m r^6 + 864 a_0^4 m r^4 w_0 +24 a_0^3 m r^2 (-12 + 7 r^2) w_0^2 + 32 a_0^2 m (1 + 9 r^2) w_0^3 -
   3 a_0 (24 m + r^2) w_0^4 + 3 w_0^5\bigg].
\end{split}
   \end{equation}

\begin{equation}\label{36}
\begin{split}
&p_{r}=\frac{1}{w_0^6} 2 a_0 \bigg[ 864 a_0^5 m r^6 - 864 a_0^4 m r^4 w_0 + 24 a_0^3 m r^2 (12 + 7 r^2) w_0^2 + 32 a_0^2 m (-1 + 2 r^2) w_0^3 +
 a_0 (88 m + 7 r^2) w_0^4 + w_0^5 \bigg].
\end{split}
   \end{equation}

   \begin{equation}\label{37}
\begin{split}
&p_{\phi}=\frac{1}{w_0^6} 2 a_0 \bigg[ 864 a_0^5 m r^6 - 864 a_0^4 m r^4 w_0 + 24 a_0^3 m r^2 (12 - 5 r^2) w_0^2 + 16 a_0^2 m (-2 + 15 r^2) w_0^3 +
 a_0 (40 m - 9 r^2) w_0^4 + w_0^5  \bigg].
\end{split}
   \end{equation}

  \begin{equation}\label{38}
\begin{split}
&p_{z}=\frac{1}{w_0^6} 2 a_0 \bigg[864 a_0^5 m r^6 - 864 a_0^4 m r^4 w_0 - 72 a_0^3 m r^2 (-4 + r^2) w_0^2 +
 32 a_0^2 m (-1 + r^2) w_0^3 + a_0 (40 m - r^2) w_0^4 + w_0^5  \bigg]
 \end{split}
   \end{equation}
\begin{figure}[h!]
\centering
\epsfig{file=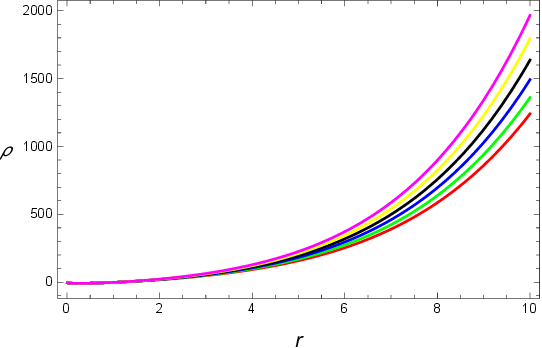, width=.32\linewidth,height=2in}
~~~~~~~~~~\epsfig{file=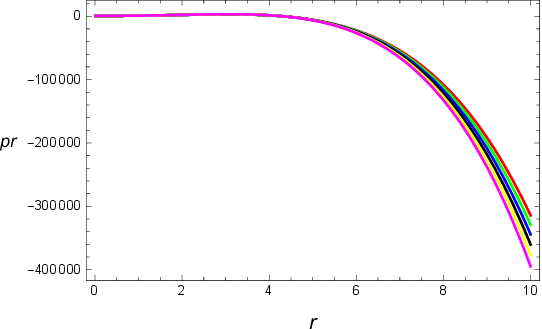, width=.32\linewidth,height=2in}\\
\epsfig{file=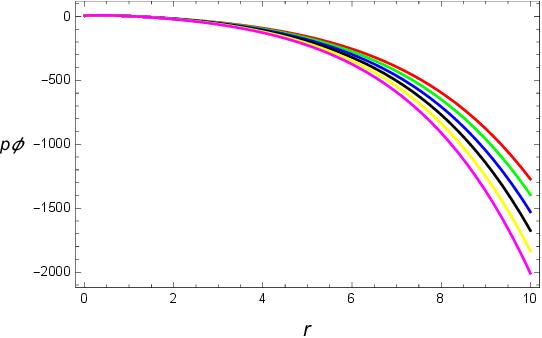, width=.32\linewidth,height=2in}
~~~~~~~~~~\epsfig{file=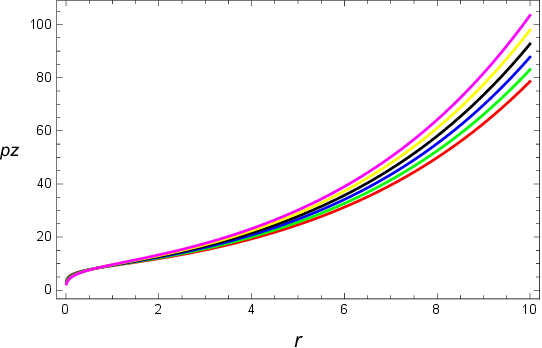, width=.32\linewidth,height=2in}
\caption{Graphical representation of energy density and pressure components for Case-II.}
\label{Fig.13}
\end{figure}

\begin{figure}[h!]
\centering
\epsfig{file=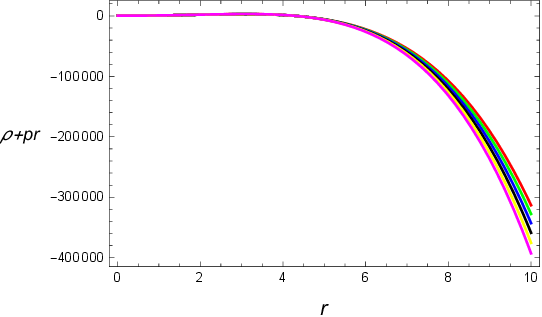, width=.32\linewidth,height=2in}
~~\epsfig{file=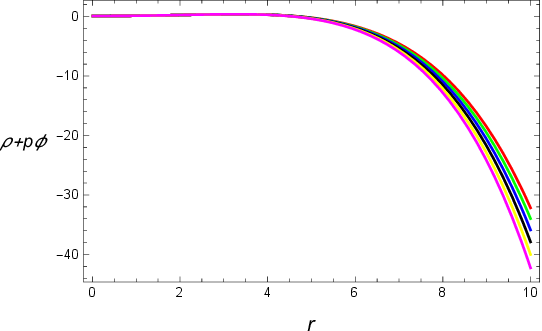, width=.32\linewidth,height=2in}
~~\epsfig{file=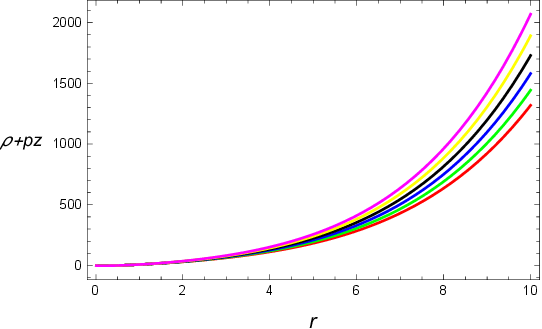, width=.32\linewidth,height=2in}
\caption{Graphical representation of NEC components of for Case-II.}
\label{Fig.14}
\end{figure}

\begin{figure}[h!]
\centering
\epsfig{file=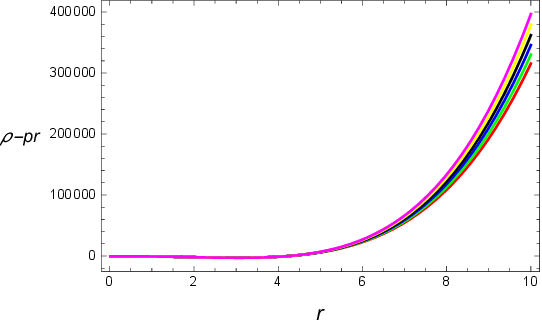, width=.32\linewidth,height=2in}
~~\epsfig{file=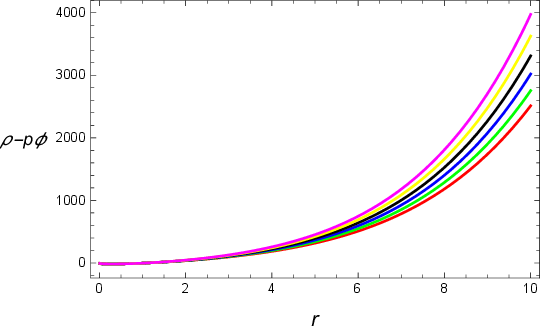, width=.32\linewidth,height=2in}
~~\epsfig{file=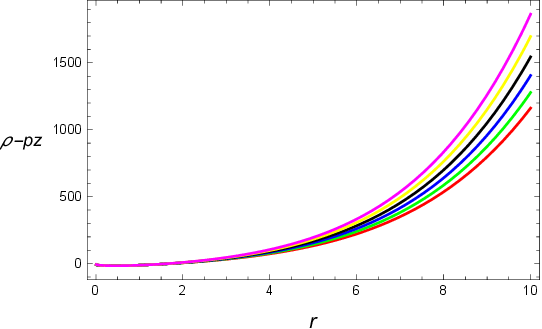, width=.32\linewidth,height=2in}
\caption{Graphical representation of DEC components for Case-II.}
\label{Fig.15}
\end{figure}

\begin{figure}[h!]
\centering
\epsfig{file=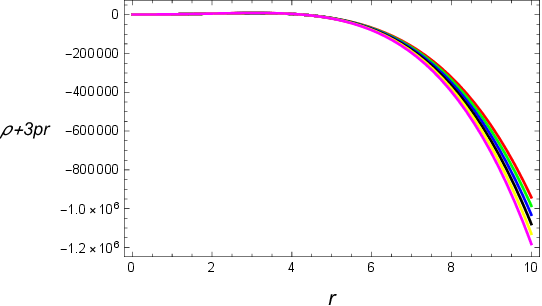, width=.32\linewidth,height=2in}
~~\epsfig{file=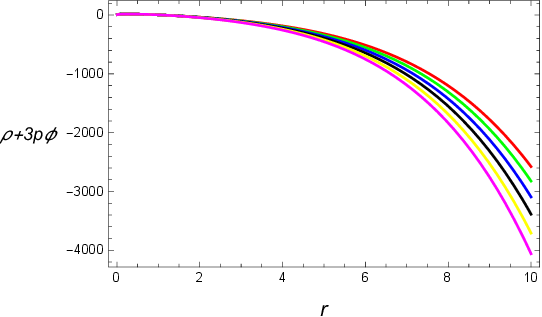, width=.32\linewidth,height=2in}
~~\epsfig{file=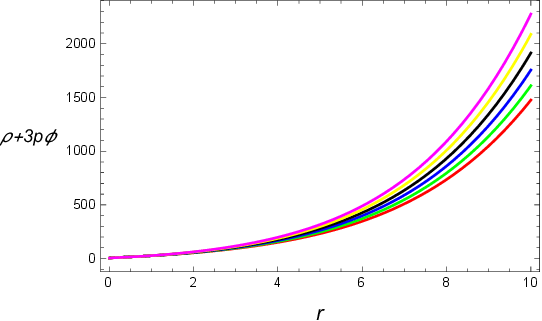, width=.32\linewidth,height=2in}
\caption{Graphical representation of DEC components for Case-II.}
\label{Fig.16}
\end{figure}

This subsection delves into a comprehensive graphical analysis of various aspects related to Case-II, covering energy density, pressure components, and energy conditions. The graphical representation of energy density and axial pressure are initially zero at the origin and becomes increasing along the radial coordinate, as depicted in Figure (\ref{Fig.13}). On the other side, the other two pressure components including radial pressure and azimuthal pressure component are opposite in nature as compared to energy density and axial pressure i.e. these components are zero at the core but it becomes negative when we move towards the boundary as seen in Figure (\ref{Fig.13}). Further scrutiny in Figure (\ref{Fig.14}) examines the sum of energy density with radial pressure and azimuthal pressure, revealing a negative and decreasing pattern. Meanwhile, the sums involving axial pressure are zero at the origin and increasing in nature as seen in Figure (\ref{Fig.14}). This observation leads to the conclusion that the null energy condition (NEC) is violated due to the negativity of $\rho+p_r$ and $\rho+p_{\phi}$. Consequently, the Weak Energy Condition (WEC), relying on the combination of energy density and NEC components, is not satisfied in this scenario. A more detailed examination of the graphical analysis for components of the dominant energy condition (DEC) unveils positive and increasing trends for the differences involving energy density with radial pressure, azimuthal pressure, and axial pressure, as observed in Figure (\ref{Fig.15}). This suggests the satisfaction of DEC for this particular case. Shifting focus to the strong energy condition (SEC), the graphical representation illustrates negativity in two components related to the sum of energy density with three times radial pressure and azimuthal pressure. However, the graphical analysis of the sum of energy density with three times axial pressure is positive, as shown in Figure (\ref{Fig.16}). Consequently, it can be summarized that SEC is violated for this case.
In summary, the graphical assessments strongly indicate a breach of null energy conditions in specific regions, hinting at the potential existence of cylindrical wormholes in those areas. These findings carry significant implications for further exploration and understanding within the realm of theoretical physics.

\subsection{Case-III}

\begin{equation}\label{39}
a(r)=a_{0}+ln[w(r)],~~~b(r)=ln[w(r)],~~~w(r)=w_{0}+w_{1}r,
 \end{equation}
 where, $a_{0}, b_{0}$ and $c_{0}$ are any arbitrary constants. Substituting these metric potentials in Eqs. (\ref{14})-(\ref{17}), we get

\begin{equation}\label{40}
\begin{split}
&\rho=w_1^2 (w_0 + r w_1)^{2 (-6 + a_0)} \bigg[(w_0 + r w_1)^8 - 8 (-1 + a_0^2)^3 m w_1^4 (w_0 + r w_1)^{4 a_0} + 4 (-1 + a_0) (1 + a_0)
\\&  (-23 + 4 a_0 (1 + a_0)) m w_1^2 (w_0 + r w_1)^{4 + 2 a_0} \bigg].
\end{split}
   \end{equation}

\begin{equation}\label{41}
\begin{split}
&p_{r}=8 (-1 + a_0^2)^3 m w_1^6 (w_0 + r w_1)^{ 6 (-2 + a_0)} + (3 + 4 a_0^2) w_1^2 (w_0 + r w_1)^{-4 + 2 a_0}4 (-1 + a_0) (1 + a_0)
\\&  (71 + 2 a_0 (-29 + 8 a_0)) m w_1^4 (w_0 +  r w_1)^{-8 + 4 a_0}.
\end{split}
   \end{equation}

   \begin{equation}\label{42}
\begin{split}
&p_{\phi}=w_1 (w_0 + r w_1)^{2 (-6 + a_0)}  \bigg[ (w_0 - (1 + a_0 + 4 a_0^2 - r) w_1) (w_0 + r w_1)^8 + 8 (-1 + a_0^2)^3 m w_1^5 (w_0 +
\\& r w_1)^{4 a_0}4 (-1 + a_0) (1 + a_0) m w_1^2 (w_0 + r w_1)^{ 4 + 2 a_0} (w_0 + (23 - 33 a_0 + 6 a_0^2 + r) w_1)  \bigg].
\end{split}
   \end{equation}

  \begin{equation}\label{43}
\begin{split}
&p_{z}=-w_1^2 (w_0 + r w_1)^{2 (-6 + a_0)} \bigg[(w_0 + r w_1)^8 - 8 (-1 + a_0^2)^3 m w_1^4 (w_0 + r w_1)^{4 a_0}  - 4 (-1 + a_0)
\\&(1 + a_0) (23 + 4 (-5 + a_0) a_0) m w_1^2 (w_0 + r w_1)^(4 + 2 a_0) \bigg]
 \end{split}
   \end{equation}

\begin{figure}[h!]
\centering
\epsfig{file=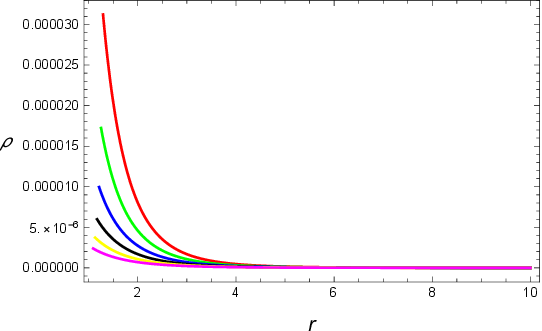, width=.32\linewidth,height=2in}
~~~~~~~~~~\epsfig{file=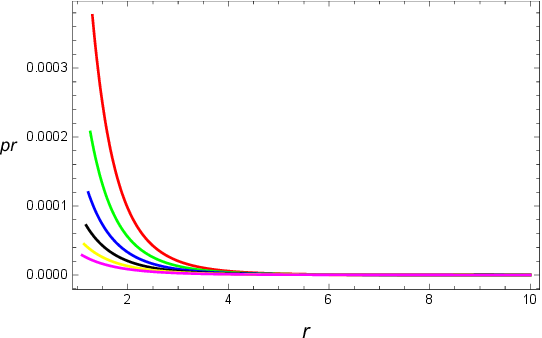, width=.32\linewidth,height=2in}\\
\epsfig{file=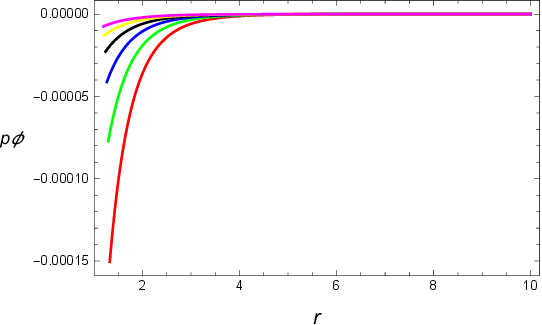, width=.32\linewidth,height=2in}
~~~~~~~~~~\epsfig{file=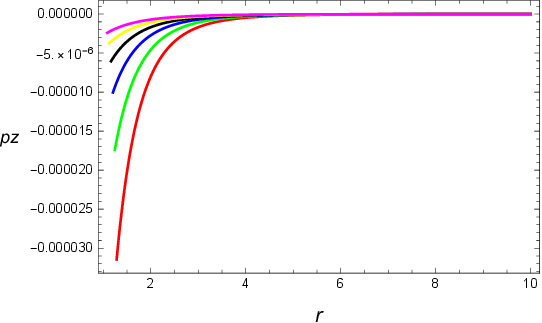, width=.32\linewidth,height=2in}
\caption{Graphical representation of energy density and pressure components for Case-III.}
\label{Fig.17}
\end{figure}

\begin{figure}[h!]
\centering
\epsfig{file=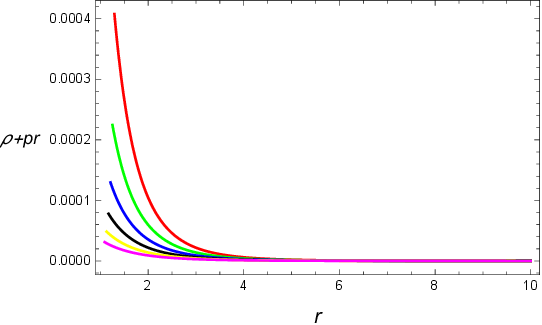, width=.32\linewidth,height=2in}
~~\epsfig{file=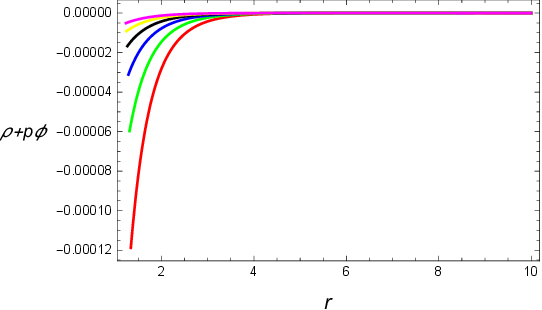, width=.32\linewidth,height=2in}
~~\epsfig{file=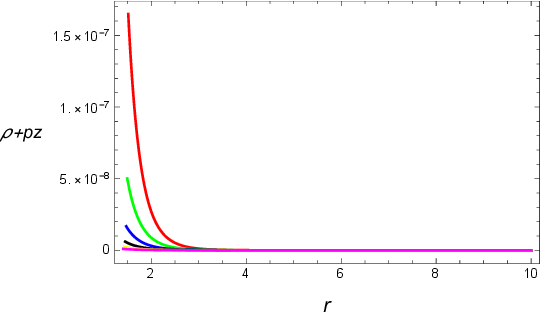, width=.32\linewidth,height=2in}
\caption{Graphical representation of NEC components of for Case-III.}
\label{Fig.18}
\end{figure}

\begin{figure}[h!]
\centering
\epsfig{file=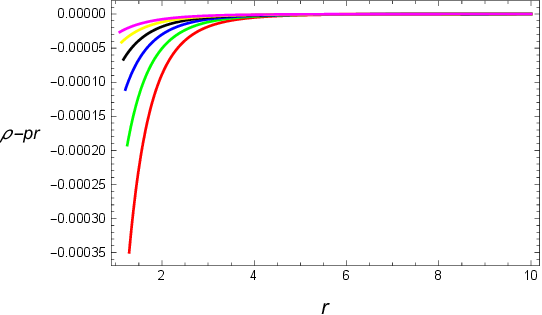, width=.32\linewidth,height=2in}
~~\epsfig{file=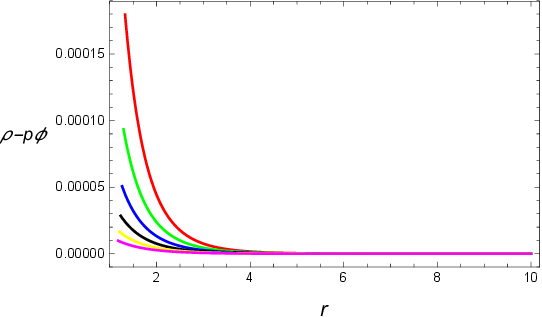, width=.32\linewidth,height=2in}
~~\epsfig{file=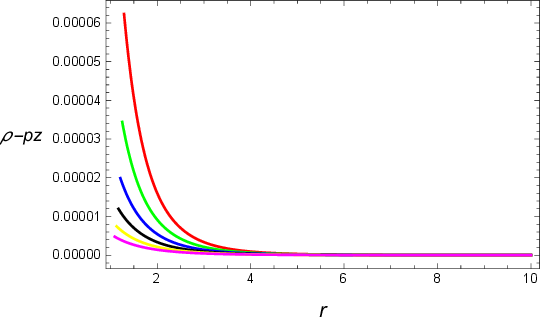, width=.32\linewidth,height=2in}
\caption{Graphical representation of DEC components for Case-III.}
\label{Fig.19}
\end{figure}

\begin{figure}[h!]
\centering
\epsfig{file=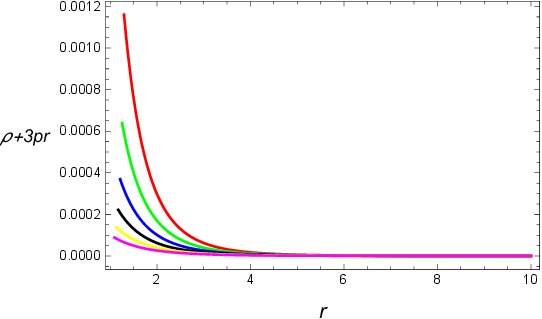, width=.32\linewidth,height=2in}
~~\epsfig{file=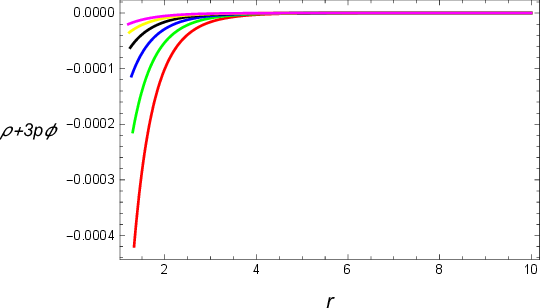, width=.32\linewidth,height=2in}
~~\epsfig{file=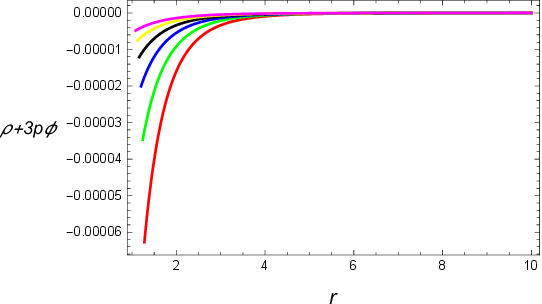, width=.32\linewidth,height=2in}
\caption{Graphical representation of DEC components for Case-III.}
\label{Fig.20}
\end{figure}

In this part, we conduct a thorough graphical analysis of different aspects related to Case-III, including energy density, pressure components, and energy conditions. As depicted in Figure (\ref{Fig.17}), the graphical representation illustrates that both energy density and radial pressure are maximum near the center but it shows decreasing nature along the radial coordinate. In contrast, the axial pressure and azimuthal pressure components follow an opposing trend, commencing from zero at the core and transitionary to negative values as we approach the boundary. Figure (\ref{Fig.18}) further explores the analysis, revealing a decreasing pattern in the sum of energy density with radial pressure and axial pressure. Moreover, the sum of energy density with the azimuthal pressure shows negative pattern as shown in Figure (\ref{Fig.18}). This finding leads to the inference that the null energy condition (NEC) is breached owing to the negativity of $\rho+p_r$ and $\rho+p_{\phi}$. Consequently, the Weak Energy Condition (WEC), which depends on the interplay of energy density and NEC components, is not met in this scenario. A more detailed scrutiny of the graphical analysis concerning the components of the dominant energy condition (DEC) uncovers positive and decreasing trends in the differences related to energy density with azimuthal pressure, and axial pressure, as evidenced in Figure (\ref{Fig.19}). On the other side, the differences related to energy density with radial pressure uncovers positive and increasing trends in the same figure. Negative behaviors of these components indicate that DEC is not satisfied for this specific case. Shifting focus to the strong energy condition (SEC), the graphical representation illustrates negativity in two components related to the sum of energy density with three times axial pressure and azimuthal pressure. However, the graphical analysis of the sum of energy density with three times radial pressure is positive, as shown in Figure (\ref{Fig.20}). Consequently, it can be summarized that SEC is violated for this case. To sum up, the graphical evaluations strongly suggest a violation of null energy conditions in certain regions, implying the possible presence of cylindrical wormholes in those areas. These discoveries hold noteworthy implications for advancing exploration and comprehension in the field of theoretical physics.


\section{Conclusion}
In this study, we have explored cylindrically symmetric solutions using the $f(R,\phi,X)$ gravity model, which proves to be a highly compatible and versatile framework. This model combines the Ricci Scalar, scalar potential, and kinetic term, enhancing the generality and appeal of our theoretical framework. Field equations have been derived in terms of density and pressure expressions, providing mathematical formulations for $\rho$, $p_{r}$, $p_{\phi}$, and $p_{z}$. Initially, we examined six distinct cases of cylindrically symmetric solutions and subsequently delved into the Levi Civita Solutions, as well as a specific case known as Cosmic string solutions. The study yields several important outcomes, summarized as follows:

\begin{itemize}
\item In summary, our graphical analysis of Levi Civita spacetime reveals intriguing patterns in energy density and pressure components. The violation of null energy conditions (NEC) and Weak Energy Condition (WEC) is evident due to negative trends in $\rho+p_r$. However, the Dominant Energy Condition (DEC) is satisfied, as observed in the positive and decreasing trends of the differences in energy density with radial, azimuthal, and axial pressure. Conversely, the violation of Strong Energy Condition (SEC) is apparent in the negative trends of the sums involving three times radial, azimuthal, and axial pressure components. This prompts the intriguing possibility of cylindrical wormholes within specific regions, emphasizing the need for further exploration and understanding in theoretical physics.
\item The detailed graphical analysis of cosmic string spacetime reveals intriguing patterns. While energy density displays a positive and decreasing trend, radial and axial pressures exhibit negative trends with increasing nature. Azimuthal pressure initially shows a negative and increasing pattern but transitions to negative behavior near the boundary. The violation of null energy conditions (NEC) and subsequent non-compliance with the Weak Energy Condition (WEC) are evident. Despite the satisfaction of Dominant Energy Condition (DEC) in certain aspects, its violation is observed for the cosmic string case in the azimuthal pressure difference. Additionally, the Strong Energy Condition (SEC) is breached, indicated by negativity in all components involving the sum of energy density with three times radial, azimuthal, and axial pressure. These findings suggest the potential presence of cylindrical wormholes in specific regions, posing significant implications for further exploration in theoretical physics.
\item The detailed graphical analysis of Case-I highlights intriguing trends in energy density and pressure components. The violation of null energy conditions (NEC) implies potential cylindrical wormholes in specific regions, challenging the satisfaction of the Weak Energy Condition (WEC). However, the Dominant Energy Condition (DEC) holds, while the Strong Energy Condition (SEC) is breached. These findings bear significant implications for further exploration in theoretical physics.
\item Regarding to the discussion of Case-II highlights distinctive trends in energy density and pressure components, revealing a violation of null energy conditions. The confirmation of SEC violation suggests potential cylindrical wormholes in specific regions. These findings hold significant implications for advancing theoretical physics exploration.

\item In summary, our analysis of Case-III unveils distinct trends in energy density and pressure components, indicating a breach of null energy conditions. The violation of both Weak and Dominant Energy Conditions and the confirmation of SEC violation suggest possible cylindrical wormholes in specific regions. These findings carry significant implications for advancing theoretical physics exploration.
\end{itemize}

\section*{Data Availability Statement}
\hskip\parindent
\small
No data was used for the research in this article. It is pure mathematics.

\section*{Conflict of Interest}
\hskip\parindent
\small
The authors declare that they have no conflict of interest.

\section*{Contributions}
\hskip\parindent
\small
 We declare that all the authors have same contributions to this paper.

\section*{Data Availability Statement}
The authors declare that the data supporting the findings of this study are available within the article.

\section*{Acknowledgement}
Adnan Malik acknowledges the Grant No. YS304023912 to support his Postdoctoral Fellowship at Zhejiang Normal University, China.

\end{document}